\documentclass[12pt]{article}
\usepackage{graphicx}
\usepackage{a4wide}

\def\pprime{p^{\,\prime}}
\def\bq{\begin{quote}}
\def\eq{\end{quote}}

\def\bq{\begin{quote}}
\def\eq{\end{quote}}

\def\bq{\begin{quote}}
\def\eq{\end{quote}}
\def\nono{\nonumber}
\def\pgg{{\gamma\gamma^\ast\to\pi^0}}
\def\bpg{{B\to\pi\ell\nu_\ell}}
\def\btp{{B\to\pi}}
\def\ptp{{\pi\to\pi\gamma^*}}
\def\dirac#1{#1\llap{/}}
\def\pv#1{\vec{#1}_\perp}
\def\as{\alpha_s}
\def\lqcd{\Lambda_\mathrm{QCD}}

\begin{document}

\newcommand{\sheptitle}
{Sudakov effects in \boldmath{$\bpg$} form factors}

\newcommand{\shepauthor}
{S. Descotes-Genon~\footnote{E-mail:\texttt{sdg@hep.phys.soton.ac.uk}}
and C.T. Sachrajda~\footnote{E-mail:\texttt{cts@hep.phys.soton.ac.uk}} }

\newcommand{\shepaddress}
{Department of Physics and Astronomy, University of Southampton,\\
Southampton, SO17 1BJ, U.K.}

\newcommand{\shepabstract}
{In order to obtain fundamental information about the Standard
Model of particle physics from experimental measurements of
exclusive hadronic two-body $B$-decays we have to be able to
quantify the non-perturbative QCD effects. Although approaches
based on the factorization of mass singularities into hadronic
distribution amplitudes and form factors provide a rigorous
theoretical framework for the evaluation of these effects in the
heavy quark limit, it is not possible to calculate the
$O(\lqcd/m_b)$ corrections in a model-independent way, because of
the presence of non-factorizing long-distance contributions. It
has been argued that Sudakov effects suppress these contributions
and render the corresponding corrections perturbatively calculable
in terms of the distribution amplitudes. In this paper we examine
this claim for the simple and related example of semileptonic
$\btp$ decays (which have similar long-distance contributions) and
conclude that it is not justified. The uncertainties in our
knowledge of the mesons' distribution amplitudes imply that the
calculations of the form factors are not sufficiently precise to
be useful phenomenologically. Moreover, it appears that a
significant fraction of the contribution comes from the
non-perturbative region of large impact parameters, and is
therefore uncalculable. We also raise a number of theoretical
issues in the derivation of the underlying formalism. Our
conclusion is therefore a disappointing one. For $B$-decays it is
not possible to invoke Sudakov effects to calculate amplitudes for
decays which have long-distance divergences (end-point
singularities) in the standard hard-scattering approach.}

\newcommand{\shepkeywords}
{PACS number(s): 13.20.He, 12.38.Bx, 12.38.Cy, 12.15.Hh}

\begin{titlepage}
\begin{flushright}
September 27 2001\\ SHEP/01-27\\
\end{flushright}
\begin{center}
{\large{\bf \sheptitle}}
\bigskip \\ \shepauthor \\ \mbox{} \\ {\it \shepaddress} \\
\vspace{.5in}
{\bf Abstract} \bigskip
 \end{center}
\shepabstract
\\
\vspace{.3in}
\\
\shepkeywords
\\
\bigskip
\setcounter{page}{0} \setcounter{footnote}{0}
\end{titlepage}

\section{Introduction}\label{sec:intro}

The study of $B$-decays is central to the development of our
understanding of the standard model of particle physics and to the
determination of its properties, parameters and limitations. The
BaBar~\cite{babar} and Belle~\cite{belle} $B$-factories as well as
other experiments will continue to provide a large amount of
experimental data on $B$-decays in general and on two-body
hadronic decays in particular. A measurement of the mixing-induced
CP-asymmetry in the golden mode $B\to J/\Psi\,K_S$ (and related
decays) for which there is a single weak phase, allows one to
determine $\sin(2\beta)$ without any hadronic uncertainties (where
$\beta$ is one of the angles of the unitarity triangle). For other
decay modes however, our inability to quantify the strong
interaction effects prevents us from being able to determine
fundamental information about the standard model from the measured
branching ratios and asymmetries with the desired precision.

Recently it was discovered that, \textit{in the $m_b\to\infty$
limit} (where $m_b$ is the mass of the $b$-quark), the
mass-singularities in two body hadronic $B$-decays
\textit{factorize}, so that the corresponding amplitudes can be
written as convolutions of universal non-perturbative quantities
(the light-cone distribution amplitudes of the mesons and
semi-leptonic form factors) and perturbatively calculable
hard-scattering kernels~\cite{BBNS}. This provides a theoretically
rigorous framework for the evaluation of decays $B\to M_1 M_2$
(where $M_1$ and $M_2$ are two mesons, at least one of which is
light) in the heavy quark limit.

The physical mass of the $b$-quark is only about 4.5\,GeV however,
and so $O(\lqcd/m_b)$ corrections are significant
(we refer to contributions which are suppressed by powers of
$1/m_b$ as power corrections). Indeed there are ``chirally
enhanced" power corrections for which the scale is not
$\lqcd$ but the much larger $m_K^2/(m_s+m_d)$. In
other cases, contributions which are suppressed by powers of
$1/m_b$ may be enhanced by CKM factors or for other reasons. We
would therefore very much like to be able to compute the power
corrections reliably. In general, mass singularities do not
factorize for the power corrections, and one is therefore reduced
to using model or phenomenological estimates or exploiting flavour
symmetries, hence losing precision and predictive power. There is
a school of thought, however, which claims that Sudakov effects
regulate these mass-singularities in such a way that they are
calculable in perturbation theory~\cite{sudakovb}. Indeed,
calculations of $B\to M_1 M_2$ decay amplitudes which include
power corrections and which use Sudakov effects to suppress the
long distance effects are being
presented~\cite{pqcd}--\cite{powercorr}.
This approach is frequently referred to as the pQCD formalism.

The possibility that the power corrections may also be calculable
in perturbation theory is an exciting one of course, although one
may have doubts whether Sudakov effects are effective at scales as
low as $m_b$. In this paper we reexamine the calculations of the
amplitudes for a related but simpler process, $B\to\pi$
semileptonic decays. In each order of perturbation theory this
process is singular at long-distances. It is for this reason that
the values of the form factors are used as a non-perturbative
input in the factorization formulae~\cite{BBNS}. On the other
hand, in the pQCD approach, the Sudakov logarithms are resummed
and the semileptonic form factors are claimed to be
calculable~\cite{Li-Sanda,Li-Yu}. In this context we address two
important questions:
\begin{enumerate}
\item Can Sudakov effects be used to evaluate the form-factors in
perturbation theory \textit{in principle}? In particular we
examine whether all (or almost all) of the contribution to the
amplitudes comes from the perturbative region.
\item Are the form-factors calculable \textit{in practice}? Specifically
we ask whether one can obtain phenomenologically useful results,
given the uncertainties in the mesons' wave functions.
\end{enumerate}
In addition we examine some of the theoretical steps in the
derivation of the pQCD formula for the semileptonic amplitudes in
terms of the Sudakov form factors.

The results of our study lead to disappointing answers to both
questions. When we apply the formulae used in the pQCD approach we
find that our ignorance of the mesons' wave functions leads to
very large uncertainties in the predictions for the form factors.
Moreover a significant fraction of the result comes from regions
of phase space where perturbation theory (and hence the formulae
which are being used) do not apply. We should stress that our
numerical results are not in disagreement with those in
refs.~\cite{pqcd}--\cite{Li-Sanda} and particularly in
ref.~\cite{Li-Yu}. The conclusions which we draw from our
calculations however, are profoundly different from those of these
authors. We believe that our calculations show that the pQCD
approach cannot be used to make reliable predictions for the form
factors, and hence also in the evaluation of the $B\to M_1M_2$
decay amplitudes and in the subsequent determination of the
parameters of the unitarity triangle and other studies of
CP-violation. This is the main conclusion of our study and our
motivation in writing this paper is to try to open a debate of
this very important issue.

We also have some reservations about the theoretical foundations
of the pQCD formula for semileptonic decays of $B$-mesons. In
particular we do not understand the framework for the derivation
of the Sudakov factor for the B-meson. We expect the Sudakov
suppression in the $B$-meson's distribution amplitude to be weak,
and soft configurations with large transverse separations to
contribute significantly to the $\bpg$ form factors. We will
explain our concerns in some detail in sections~\ref{sec:bdas} and
\ref{sec:numerical}.

The plan of the remainder of this paper is as follows. In
section~\ref{sec:btpstd} we recall why the $\btp$ form factors
cannot be computed using the standard factorization approach. In
section~\ref{sec:pqcd}, we present an outline of the framework
used in the pQCD approach. The use of Sudakov suppression of long
distance contributions has been introduced into hard exclusive
processes involving light hadrons in order to improve the
precision of the calculations, and is now being used also in
$B$-physics. In the following three sections we investigate the
ingredients of the pQCD formalism for semileptonic $B$-decays in
detail and highlight our concerns. Section~\ref{sec:sudakovpionda}
contains a discussion of the distribution amplitude of the pion
and section~\ref{sec:bdas} is devoted to a critical examination of
the derivation of the Sudakov factor for the distribution
amplitude(s) of the $B$-meson. We also investigate whether the
approximation of using a single distribution amplitude for the
$B$-meson is a valid one. In section~\ref{sec:numerical}, the pQCD
expressions for the $\btp$ form factors are shown to be very
sensitive to the shape of the mesons' distribution amplitudes, and
to contain significant uncalculable long-distance contributions.
Section~\ref{sec:concs} contains a summary of our principal
conclusions. In the appendix we introduce a model distribution
amplitude for the $B$-meson, which satisfies the constraints due
to the equations of motion in longitudinal and transverse momentum space.

\section{$\btp$ form factors in the factorization approach}
\label{sec:btpstd}
In this section we consider the semi-leptonic decay $B(p)\to
\pi(p')\ell \nu_\ell$ in the standard factorization approach,
introduced by Brodsky and Lepage and by Efremov and Radyushkin
(BLER)~\cite{Brod-Lep,Chern}. We introduce the main ingredients of
this approach, and in particular the light-cone distribution
amplitudes which describe the parton content of the mesons
relevant for this process. We then recall why long-distance
singularities prevent the computation of $\btp$ form factors in
the standard approach~\cite{Chern-Zhit}.

\begin{figure}[t]
\begin{center}
\includegraphics[height=5cm]{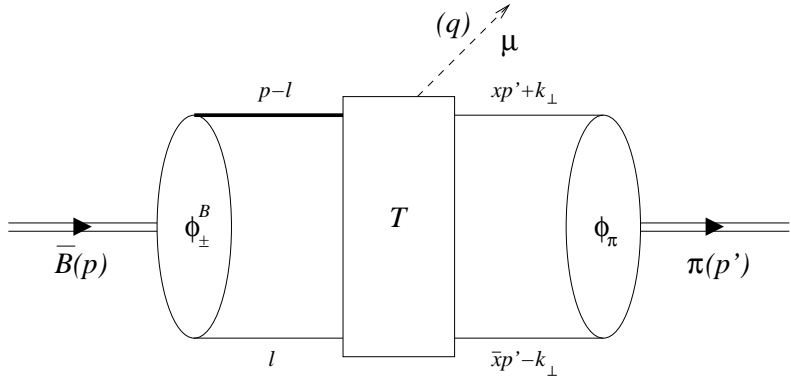}
\caption{{\footnotesize Diagrammatic representation of the $\btp$
semileptonic decay in the factorization approach. The thick line
represents a $b$-quark.}} \label{fig:kinem2}
\end{center}
\end{figure}

We start by defining the kinematics and our notation. The
amplitude for this decay can be written in the form:
\begin{equation} {\mathcal A}(p,p')=\frac{G_F}{\sqrt{2}}\,V_{ub}\
   \left(\bar{\nu}_\ell\gamma_\mu(1-\gamma_5)\ell\right)\
   \langle\,\pi(p^{\,\prime})\,|\,\bar{u}\gamma_\mu b\,|\,\bar{B}(p)\,\rangle\ ,
\end{equation}
where $l$ and $\bar\nu_l$ represent the wave functions of the
charged lepton and neutrino respectively and $\bar u$ and $b$ are
quark fields.
The non-perturbative QCD effects are contained in the hadronic
matrix element $\langle\,\pi(p^{\,\prime})\,|\,\bar{u}\gamma_\mu
b\,|\,\bar{B}(p)\,\rangle$, which can be parametrized in terms of
two invariant form factors. We choose to use a conventional
definition of the form factors (in a helicity basis):
\begin{equation}
\langle\,\pi(p^{\,\prime})\,|\,\bar{u}\gamma_\mu b\,|\,\bar{B}(p)\,\rangle
  =F_+(q^2) (p+p')_\mu +
  (M_B^2-M_\pi^2)\frac{F_0(q^2)-F_+(q^2)}{q^2} q_\mu\ ,
\label{eq:defformfac}
\end{equation}
where $q_\mu=p_\mu-p'_\mu$. At $q^2=0$ we have $F_0(0)=F_+(0)$.
The notation is illustrated in Fig.~\ref{fig:kinem2}.

We work in the rest frame of the $B$-meson, and neglect the mass
of the pion. We introduce light-cone coordinates, writing
$k=(k_+/\sqrt{2},k_-/\sqrt{2},k_\perp)$ with $k_{\pm}\equiv(k_0\pm k_3)$,
and with the metric for the transverse components chosen such that
$(k_\perp)^\mu (l_\perp)_\mu=-\pv{k}\cdot \pv{l}$. It will also be
convenient to introduce two light-like vectors
$n_+\equiv(\sqrt{2},0,\pv{0})$ and $n_-\equiv(0,\sqrt{2},\pv{0})$.

Using light-cone coordinates,
$p=(M_B/\sqrt{2},M_B/\sqrt{2},\pv{0})$ and we chose the momentum
of the pion to be in the $-$ direction, $p'=(0,\eta
M_B/\sqrt{2},\pv{0})$. In terms of $\eta$, the energy of the pion
is given by $p^{\,\prime\,0}=\eta M_B/2$ and the invariant mass of
the lepton pair by $m^2_{\ell\nu}=(1-\eta)M^2_B$. The physical
range for $\eta$ is therefore given by $0\leq \eta \leq 1$.

In the standard hard-scattering approach, as represented in
Fig.~\ref{fig:kinem2}, the form factors are obtained from
convolutions of the mesons' distribution amplitudes ($\phi_\pi$
and $\phi^B_\pm$) and the perturbative kernel $T$. In the
following subsection we discuss the distribution amplitudes before
returning to the convolution itself in section~\ref{subsec:hsaff}.

\subsection{Meson distribution amplitudes}\label{subsec:mesonda}

In order to evaluate the decay amplitude, we have to perform the
convolution illustrated in fig.~\ref{fig:kinem2}. In this
subsection we discuss one of the principal ingredients of this
convolution, namely the distribution amplitudes (or DAs) of the
incoming and outgoing mesons. We will consider only the
leading-twist amplitudes here.

The (leading-twist) distribution amplitude of the pion is defined as
\cite{Brod-Lep}:
\begin{equation} \label{eq:pida}
\langle \pi(p')|\bar{q}_\alpha(y) q'_\beta(z)|0\rangle
 =i \frac{f_\pi}{4} ({\dirac{p}}^\prime\gamma_5)_{\beta\alpha}
  \int_0^1 dx\,e^{i(xp'\cdot y+\bar{x}p' \cdot z)}\phi_\pi(x;\mu)\ ,
\end{equation}
where $(y-z)^2=0$, $\bar{x}=1-x$, and we choose a normalization
such that $f_\pi\simeq 131$\,MeV. A path-ordered exponential is
implicit on the left-hand side, in order to maintain gauge
invariance. The parameter $\mu$ is a renormalization scale of the
light-cone operators on the left-hand side.

The distribution amplitude is symmetric with respect to the
interchange $x\leftrightarrow \bar{x}$, and is normalized such
that $\int_0^1 dx\, \phi_\pi(x;\mu)=1$. As the renormalization
scale $\mu$ tends to infinity, $\phi_\pi$ tends to the asymptotic
limit $\phi_\pi^\mathrm{as}(x)=6x(1-x)$. For finite values of the
renormalization scale, $\phi_\pi$ can be expanded on the basis of
Gegenbauer polynomials:
\begin{equation} \label{eq:gegenexpand}
\phi_\pi(x;\mu)=6x(1-x)\left[1+\sum_{p=1}^\infty
   \alpha_{2p}(\mu) C_{2p}^{(3/2)}(2x-1)\right]\,.
\end{equation}
Only even polynomials contribute because of the symmetry
$x\leftrightarrow \bar{x}$. The lowest Gegenbauer polynomials
contributing to the sum are $C_2^{(3/2)}(u)=3/2\cdot(5u^2-1)$ and
$C_4^{(3/2)}(u)=15/8\cdot(21u^4-14u^2+1)$. The moments
$\alpha_{n}$ are multiplicatively renormalized, with the anomalous
dimensions increasing as $n$ increases~\cite{Brod-evol} (so that
the corresponding $\alpha_{2p}$ decreases more quickly with
increasing $p$ as $\mu\to\infty$).

For the distribution amplitudes of the $B$-meson, we will use the
notation and formalism developed in refs.~\cite{Groz-Neub} and
\cite{Ben-Fel} which we now briefly summarize~\footnote{We note
the reservations concerning the use of such distribution
amplitudes for heavy mesons at next-to-leading order in the
perturbative expansion expressed in
refs.~\cite{korchemsky,gkregensburg}. At this order, which is
beyond the scope of our study, it is likely that the formalism
will have to be modified, at least to include the dependence on
the transverse momentum explicitly.}. In the heavy-quark limit,
the most general decomposition of the light-cone matrix element
for the $B$-meson is \cite{Groz-Neub,Ben-Fel}:
\begin{equation} \langle\, 0|\bar{q}_\beta(z)\, b_\alpha(0)|\bar{B}(p)\rangle
  =-\frac{if_B}{4}\left[\frac{\dirac{p}+m}{2}
     \left\{2\tilde\phi_+^B(t)
     +\frac{\tilde\phi_-^B(t)-\tilde\phi_+^B(t)}{t}\dirac{z}\right\}
     \gamma_5\right]_{\alpha\beta}\ , \label{eq:matelemb}
\end{equation}
where $z^2=0$, $v=p/m$, $t=v\cdot z$ and a path-ordered
exponential is implicit on the left-hand side again. We are
working only to leading twist order and therefore have set
$M_B=m_b\equiv m$. Recalling that $\langle\,
0|\bar{q}(0)\gamma_\mu\gamma_5 b(0)|\bar{B}(p)\rangle=if_B p^\mu$,
the choice of prefactor in eq.~(\ref{eq:matelemb}) implies that
\begin{equation}
\tilde\phi_+^B(t=0)=\tilde\phi_-^B(t=0)=1\ .
\label{eq:bnorm}
\end{equation}
Thus for the $B$-meson we have two distribution amplitudes,
$\tilde\phi^B_{\pm}$, to consider.

In order to evaluate the amplitudes for $\bpg$ decays we perform a
convolution of the distribution amplitudes of the $B$-meson with
the hard-scattering amplitude, as illustrated in
Fig.~\ref{fig:kinem2}. Let $l$ denote the momentum of the
light quark in the $B$-meson and recall that we have chosen the
momentum of the outgoing pion to be in the $-$ direction. The
$O(\as)$ hard-scattering amplitude $T$ is then independent of $l_-$.
We are thus allowed to set $z_+=0$ in the decomposition in
eq.~(\ref{eq:matelemb}). Following ref.~\cite{Ben-Fel} it is
possible to write the convolution in the form $\int_0^\infty
dl_+\,\{P^B T(l)\}$, with $l$ set to $(\omega/\sqrt{2},0,\pv{0})$
once the momentum-space projection operator $P^B$ has been
applied. The projection operator is equal to:
\begin{equation}
P^B_{\beta\alpha}=-\frac{if_B}{4}
  \left[\frac{\dirac{p}+m}{2}\left\{
  \phi_+^B(\omega) \dirac{n}_+ + \phi_-^B(\omega) \dirac{n}_-
  -\Delta(\omega)\gamma^\mu \frac{\partial}{\partial l_\perp^\mu}
  \right\}\gamma_5\right]_{\alpha\beta}\ , \label{eq:bmesonproj}
\end{equation}
with $\Delta(\omega)\equiv\int_0^\omega d\ell \,(\phi^B_-(\ell) -
\phi^B_+(\ell)\,)$. The distribution amplitudes on the right-hand
side of eq.~(\ref{eq:bmesonproj}) are written in momentum space,
\begin{equation}
\phi^B_\pm(\omega)=\frac{1}{2\pi} \int_{-\infty}^\infty
   dt\, e^{i\omega t}\,\tilde\phi_\pm^B(t)\ .
\end{equation}
Since the components of the light quark in the $B$-meson are
expected to remain of order $\lqcd$ or less, the distribution
amplitudes will be suppressed for $\omega$ larger than $\lqcd$.
Note also that eq.~(\ref{eq:bnorm}) implies that the distribution
amplitudes in momentum space are normalized such that $\int
d\omega\, \phi_+^B(\omega)=\int d\omega\, \phi_-^B(\omega)=1$.

For a precise calculation of the form factors it is clearly
important to have as much information about the mesons'
distribution amplitudes as possible. Of particular importance is
the behaviour of the distribution amplitudes at the end-point
$l_+\to 0$. To gain some insight into this behaviour,
we follow ref.~\cite{Groz-Neub} and note that the leading-twist
decomposition in eq.~(\ref{eq:matelemb}) leads to expressions for
the distribution amplitudes in terms of matrix elements of the
pseudoscalar density and axial current:
\begin{eqnarray}
\langle 0\,|\,\bar{q}(z)\gamma_5 b(0)\,|\bar{B}(p)\rangle &=&
  -if_BM_B \tilde\phi_P^B\ ,\ \ \ \textrm{and}
   \label{eq:pseudomatelem}\\
\langle 0\,|\,\bar{q}(z)\gamma_\mu\gamma_5
b(0)\,|\bar{B}(p)\rangle
  &=& f_B\left[i\tilde\phi^B_{A1} p^\mu - M_B \tilde\phi^B_{A2}
  z^\mu\right]\ ,
   \label{eq:axialmatelem}
\end{eqnarray}
where $\tilde\phi^B_P$, $\tilde\phi^B_{A_1}$ and
$\tilde\phi^B_{A_2}$ are given in terms of the distribution
amplitudes $\tilde\phi^B_\pm$ by
\begin{equation}
\tilde\phi^B_P\equiv\frac{\tilde\phi^B_++\tilde\phi^B_-}{2}\ ,
\qquad \tilde\phi^B_{A1}\equiv\tilde\phi^B_+
\qquad\textrm{and}\qquad
\tilde\phi^B_{A2}\equiv\frac{i}{2}\frac{\tilde\phi^B_+ -
\tilde\phi^B_-}{t}\,. \end{equation}
For a light pseudoscalar meson, the end-point behaviour of the
eigenfunctions of the evolution equations, together with sum-rule
inspired arguments, suggest that $\phi^B_{A1}(\omega)\sim \omega$
and $\phi^B_P(\omega)\sim 1$ for $\omega\to 0$ \cite{Chern}. If
this behaviour is also valid for the $B$-meson, we might expect
that $\phi_+^B(l_+) =O(l_+)$ and $\phi_-^B(l_+) =O(1) $ as $l_+\to 0$
\cite{Groz-Neub}. However, in spite of this argument, when
studying the validity of perturbative estimates of the form
factors in terms of the distribution amplitudes it must be
remembered that such an end-point behaviour is a conjecture.

Another theoretically motivated constraint is a consequence of the
equations of motion. Since we neglect the 3-particle distribution
amplitudes (and contributions from higher Fock states), the
equations of motion lead to the constraint~\cite{Ben-Fel}:
\begin{equation}
\phi_+^B(l_+)=-l_+ \frac{d\phi_-^B}{dl_+}(l_+)\ .
\label{eq:eom}\end{equation}
Eq.~(\ref{eq:eom}) is useful in constraining models for the
distribution amplitudes, and supports the conjectured behaviour of
the distribution amplitudes as $l_+\to 0$. We can indeed derive
the expected behaviour, that $\phi_-^B(l_+)=O(1)$ as $l_+\to 0$
and that $\phi_+^B(0)=0$, if we assume that the equation of motion
(\ref{eq:eom}) is fulfilled and that $\lambda_B^{-1}\equiv
\int_0^\infty\! dl_+\ \phi_+^B(l_+)/l_+$ is finite and
nonvanishing~\footnote{We thank Martin~Beneke for raising this
point.}. We can use the equation of motion to see that
$\lambda_B^{-1}=\phi^B_-(0)$ \cite{Ben-Fel2}, which leads to
$\phi^B_-(l_+)=O(1)$ for small $l_+$. Eq.~(\ref{eq:eom}) implies
then that $\phi^B_+(l_+)$ vanishes at the end-point.

We end this subsection by introducing two models for the
distribution amplitudes which satisfy the above constraints, and
which will be useful in our investigation of the validity and
reliability of the frameworks used to evaluate the form factors.
The first model was proposed in ref.~\cite{Groz-Neub} on the basis
of a QCD sum rule analysis:
\begin{equation}
\phi_+^B(l_+)=\frac{l_+}{\lambda^2}\exp\left[-\frac{l_+}{\lambda}\right],
\qquad
\phi_-^B(l_+)=\frac{1}{\lambda}\exp\left[-\frac{l_+}{\lambda}\right]\
.
\end{equation}
$\lambda$ is a parameter of order $\lqcd$. Another model
satisfying the above constraints is discussed in
Appendix~\ref{appsepmod}:
\begin{equation}
\phi_+^B(l_+)=\sqrt{\frac{2}{\pi \lambda^2}}
     \frac{l_+^2}{\lambda^2}\exp\left[-\frac{l_+^2}{2\lambda^2}\right]\ ,
\qquad \phi_-^B(l_+)=\sqrt{\frac{2}{\pi
\lambda^2}}\exp\left[-\frac{l_+^2}{2\lambda^2}\right].
\end{equation}
In this model, as discussed in the appendix, $\lambda$ is related
to the distribution of transverse momenta.

\subsection{Expressions for the $\btp$ form factors}
\label{subsec:hsaff}

\begin{figure}[t]
\begin{center}
\includegraphics[width=9cm]{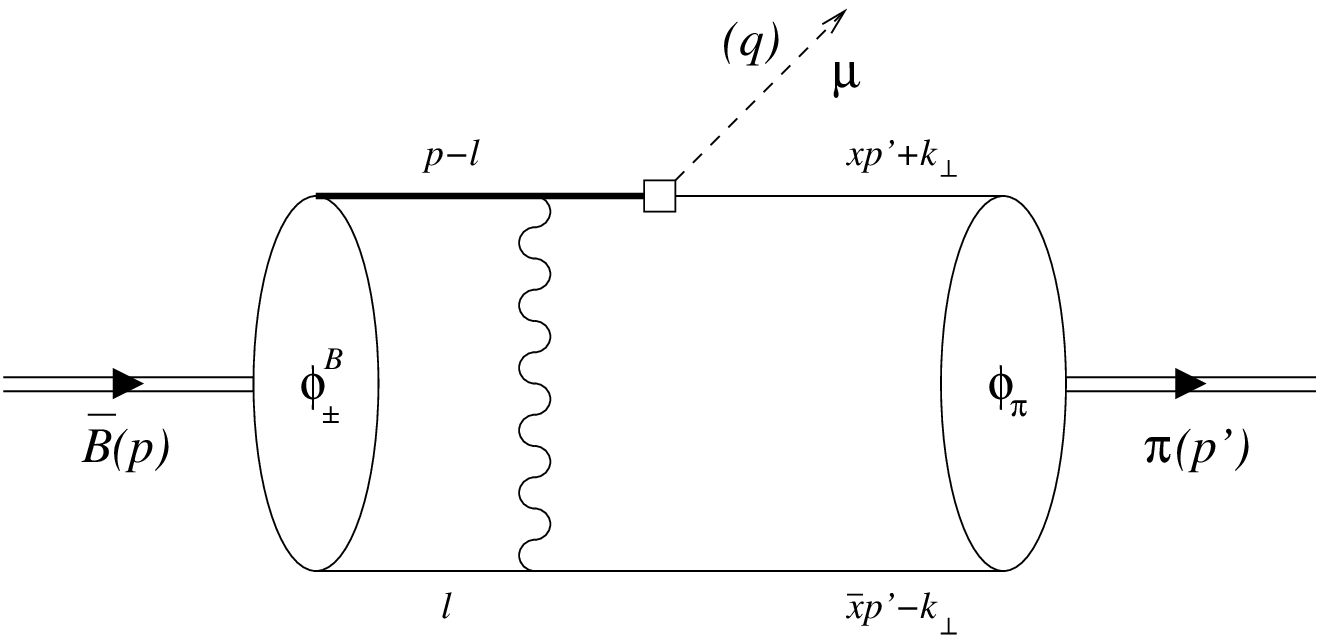}

\vspace{0.5cm}

\includegraphics[width=9cm]{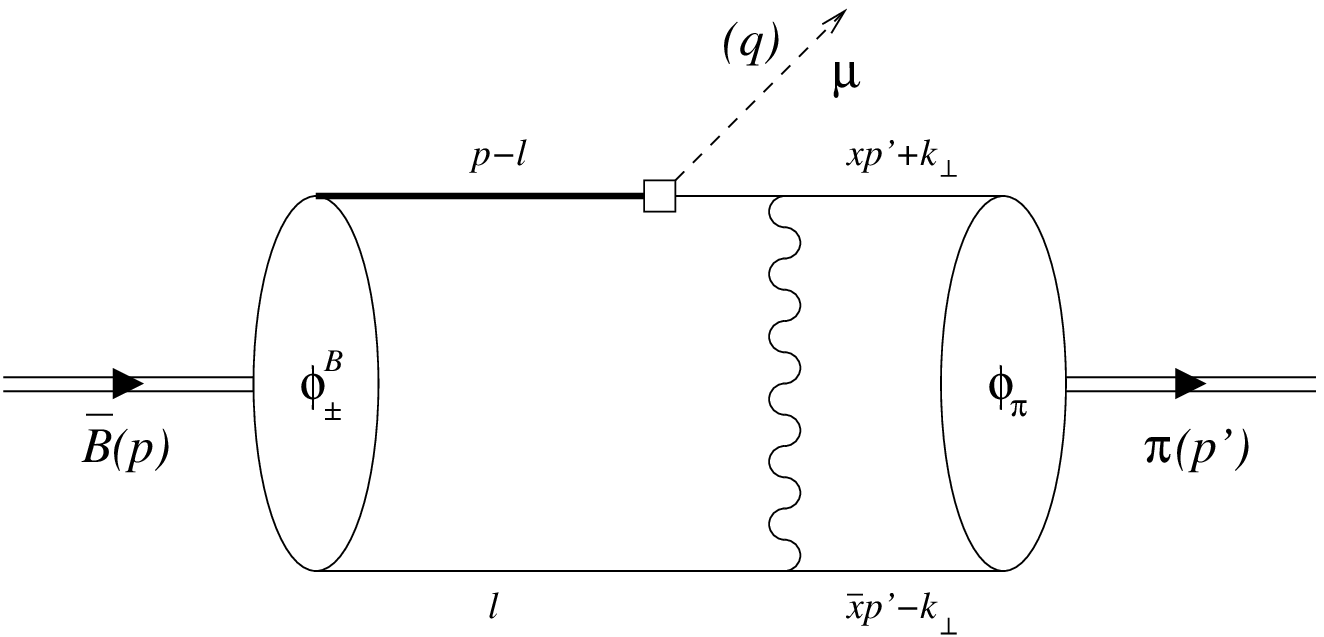}
\caption{{\footnotesize Diagrammatic representation of the
lowest-order hard-scattering kernel for $\bpg$. The transverse
momenta are introduced for the discussion in
sections~\ref{sec:pqcd} and \ref{sec:sudakovpionda}.}}
\label{btopilowest}
\end{center}
\end{figure}

We are now ready to discuss the semileptonic $\btp$ form-factors
in the standard hard-scattering approach. The leading-order
$O(\as)$ contribution to the $\btp$ form factors is given by the
two diagrams in Fig.~\ref{btopilowest} and leads to the
expressions:
\begin{eqnarray}
F_0 &=& f_\pi f_B\frac{\as\pi C_F}{N_c}
\int_0^1 dx\int dl_+
  \frac{\phi_\pi[(1+\bar{x})\phi_-^B+\bar{x}\phi_+^B]}
      {\eta m l_+ \bar{x}^2}\ ,
  \label{eq:f0std}\\
F_+ &=& f_\pi f_B\frac{\as\pi C_F}{N_c}
\int_0^1 dx\int dl_+
  \frac{\phi_\pi[(1+\bar{x})\phi_-^B+\bar{x}(2\eta-1)\phi_+^B]}
     {\eta^2 m l_+ \bar{x}^2}\ .
  \label{eq:fpstd}
\end{eqnarray}
The expression for $F_+$ appears in ref.~\cite{Ben-Fel}. In
deriving these expressions we have used the constraint from the
equation of motion eq.~(\ref{eq:eom}) to write
$\Delta(\omega)\equiv \int_0^\omega d\ell \,(\phi^B_-(\ell) -
\phi^B_+(\ell)\,)=\omega\phi_-^B(\omega)$.

From equations~(\ref{eq:f0std}) and (\ref{eq:fpstd}) we see that
the $\btp$ form factors are very sensitive to the end-point
behaviour of the distribution amplitudes for the pion and the $B$
meson. If we suppose, for example, that $\phi_\pi(x)\sim \bar{x}$
for small $\bar{x}$ and $\phi^B_-(l_+)\sim 1$ for small $l_+$, the
integrations over $x$ and $l_+$ are both divergent. It is
therefore not possible to compute $\btp$ form factors in the
standard BLER approach. It is for this reason that
the authors of ref.~\cite{BBNS} proposed to consider these form
factors as non-perturbative inputs (obtained for example from
experiment or lattice QCD) when evaluating two-body hadronic
$B$-Decays. In the pQCD approach on the other hand, Sudakov
effects are invoked to regulate the integrals at the end-points.
We now proceed to discuss this approach.

\section{Outline of the pQCD approach}
\label{sec:pqcd}

A different approach (modified factorization or pQCD) has been
introduced with the aim of solving the problem discussed in
section \ref{subsec:hsaff} above. In
refs.~\cite{pqcd}--\cite{Li-Yu} it is claimed that the
introduction of transverse momenta and the subsequent resummation
of large logarithms into Sudakov form factors regulates the
end-point singularities sufficiently to enable the $\btp$
form-factors to be computable. In this approach, the form-factors
are expressed as \cite{Li-Yu}:
\begin{equation} \label{eq:btpconvol}
F_{0,+}=\int_0^1 \!\!dx\int_0^\infty \!\!dl_+ \int
  \!\!d^2 \pv{b}\,d^2\pv{c}\,
  \tilde\Psi^B(l_+,\pv{c};\mu)\,
  \tilde{H}_{0,+}(x,l_+,\pv{b},\pv{c},\eta,m;\mu)\,
  \tilde\Psi_\pi(x,\pv{b};\mu)\ ,
\end{equation}
where:
\begin{itemize}
\item the meson distribution amplitudes $\tilde\Psi_\pi$ and
$\tilde\Psi^B$ contain the resummed Sudakov effects;
\item
$\pv{b}$ and $\pv{c}$ are the variables conjugate to the
transverse momenta of the valence quarks and correspond to their
transverse separations;
\item
$\tilde{H}_{0,+}$ are the Fourier transforms of the corresponding
hard-scattering kernels.
\end{itemize}

In the following sections we will examine the ingredients in
eq.~(\ref{eq:btpconvol}) in some detail. For presentational
purposes however, we hope that it will be instructive if we begin
with a brief discussion of the philosophy of the pQCD approach in
the simpler and much-analyzed case of the pion electromagnetic
form factor~\cite{Li-Ster,Piff,Sudacorrect}. This form factor can be computed in
the standard approach~\cite{Brod-Lep,Croatia},
leading to a convolution of
a hard-scattering kernel $T$ with the DAs of the two pions:
\begin{equation} \label{eq:pielectromagff}
F^{\ptp}(Q)=\int_0^1 du\, dv\,
  \phi_\pi(u;\mu)\, T(u,v,Q;\mu)\, \phi_\pi(v;\mu)\ ,
\end{equation}
where $Q$ is the momentum transfer. Although the convolution in
eq.~(\ref{eq:pielectromagff}) is finite, it has been
argued~\cite{Isgur,Radyushkin} that substantial contributions may
come from end-point configurations ($u,v\to 0,1$), where one or
more propagators of the hard-scattering kernel T gets close to the
mass-shell, enhancing the soft contributions. It is likely that
very large momentum transfers may have to be reached in order to
ensure a sufficient suppression of these soft contributions.

The authors of ref.~\cite{Li-Ster} proposed to exploit the ideas
developed in refs.~\cite{Collins-Soper,Botts-Ster} in order to
extend the domain of applicability of
eq.~(\ref{eq:pielectromagff}). The essential idea is to keep the
transverse momenta of the valence quarks writing:
\begin{equation} \label{eq:modifconvol}
F^{\ptp}=\int_0^1 du\, dv \int d^2 \pv{k}\,d^2\pv{l}\,
  \Psi_\pi(u,\pv{k};\mu)\,
  H(u,v,\pv{k},\pv{l},Q;\mu)\, \Psi_\pi(v,\pv{l};\mu)\ .
\label{eq:emffmom}\end{equation}
Eq.~(\ref{eq:emffmom}) is analogous to the Fourier transform of
eq.~(\ref{eq:btpconvol}). The tree-level contribution to the
hard-scattering kernel is now modified. A typical modification is
$1/(uv Q^2) \to 1/[uv Q^2+(\pv{k}+\pv{l})^2]$, thus using the
transverse momenta to regulate the end-point singularities. Since
the transverse momenta are soft, further arguments have to be
invoked in order to demonstrate that the procedure is a consistent
one.

The modified hard-scattering kernel is now convoluted with a
light-cone wave-functions which depend on both longitudinal and
transverse components of momentum. Such wave-functions are defined
by \cite{Botts-Ster}:
\begin{equation} i f_\pi p^\prime_\mu \Psi_\pi(x,\pv{k})=
  \int \frac{dz_-\,d^2\pv{z}}{(2\pi)^3}
e^{i(xp^\prime\cdot z_- - \pv{k}\cdot\pv{z})}
  \langle 0 | \bar{q}(z) \gamma_\mu \gamma_5 q'(0) | \pi(p^\prime) \rangle
  \big|_{z_+=0}
\label{eq:modifwavefn}
\end{equation}
in an axial gauge $n\cdot A=0$ [$n^2\neq 0$]. We emphasize that no
path-ordered exponential is present in the matrix element of
eq.~(\ref{eq:modifwavefn}); $\Psi_\pi$ is a gauge-dependent
quantity (indeed, as explained below, the gauge dependence is
exploited to simplify the calculation).

It is argued that gluon exchanges suppress configurations where
the quark-antiquark pair has a large transverse separation. Soft
and collinear gluon exchanges between the valence quarks lead to
large double-logarithmic effects, which need to be resummed. This
resummation is performed in impact parameter space and so we
define the Fourier conjugate of the wave-function,
\begin{equation}
\tilde\Psi_\pi(x,\pv{b})=\int
d^2 \pv{k} e^{i \pv{k}\cdot \pv{b}} \Psi(x,\pv{k})\ .
\end{equation}
Expression (\ref{eq:modifconvol}) for the form factor can be
re-expressed as an integral over the impact parameters,
\begin{equation} \label{eq:fourierconvol}
F^{\ptp}=\int_0^1 du\, dv \int d^2 \pv{b}\,d^2\pv{c}\,
  \tilde\Psi_\pi(u,\pv{b};\mu)\,
  \tilde{H}(u,v,\pv{b},\pv{c},Q;\mu)\, \tilde\Psi_\pi(v,\pv{c};\mu)
  \ .
\end{equation}
An analysis of the double logarithms from higher-orders of
perturbation theory shows that they can be resummed and included
by multiplying the wave-function in impact parameter space by a
Sudakov factor~\cite{Botts-Ster}. In the leading-logarithm
approximation
\begin{equation}
\tilde\Psi_\pi(x,\pv{b})\to
\tilde\Psi^{LL}_\pi(x,b_\perp,Q;\mu)=\exp[-{\mathcal
S}^{LL}(b_\perp,Q)]\phi_\pi(x;\mu=1/b_\perp)\ ,
\label{eq:llpida}\end{equation}
where
\begin{equation}
{\mathcal S}^{LL}(b_\perp,Q)=\frac{\as C_F}{4\pi} \ln^2(b_\perp Q)\ .
\label{eq:llsuda}
\end{equation}

When subleading logarithms are included, ${\mathcal S}$ becomes a
function of $\log[xQ/\lqcd]$ and $\log[1/(b_\perp\lqcd)]$ and
diverges when either of these logarithms vanish (see
eq.~(\ref{eq:subr}) in the
following section). This factor suppresses two regions: large
transverse separations $b_\perp$ ($b_\perp\sim O(1/\lqcd)$) and
small fractions of longitudinal momentum $x$ ($x\sim O(\lqcd/Q)$).
For the electromagnetic form factor of the pion~\cite{Li-Ster},
the Sudakov factor is considered only when it suppresses the
configurations with a large transverse separation, and it is set
to 1 outside of the region of large $b_\perp$. Potential
suppression for end-point configurations (small $x$), and (small)
enhancements for intermediate $x$ and $b_\perp$ have been
neglected.

What is generally taken for the scale of the coupling constant in
the hard-scattering kernel $\tilde{H}$ in
eq.~(\ref{eq:fourierconvol}) is the largest available virtuality
for the hard gluon, i.e. the maximum of $\sqrt{uv}Q$
(longitudinal), $1/b_\perp$ and $1/c_\perp$ (transverse). It is
then argued that the coupling constant remains small and the
perturbative approach is consistent, unless $b_\perp$ and
$c_\perp$ are both $O(1/\lqcd)$ and $u$ or $v \to
0$. Moreover, although the coupling constant is large in this soft
region, the corresponding contribution is suppressed by Sudakov
factors of the distribution amplitudes. The contribution from the
soft region is therefore argued to be small.

Before returning to the semileptonic $\btp$ decay we briefly
summarize the preceding paragraphs about the possibility that the
modified factorization approach might improve the BLER approach.
Its principal aim is to treat the end-point regions of phase
space, in order to improve the accuracy of perturbative
calculations. The singularities of the propagators are smoothed by
considering transverse components of the momenta. The scale of the
hard-scattering kernel is taken to be the maximal (longitudinal or
transverse) virtuality of the exchanged gluons. The only region of
large $\as$ corresponds to small longitudinal momenta and large
transverse separations. This region is expected to be strongly
suppressed by the Sudakov exponential and it is therefore argued
that a consistent perturbative computation is possible.

The authors of refs.~\cite{Li-Sanda,Li-Yu} claim that a similar
procedure can be followed to compute $\btp$ form factors within
the pQCD approach. Let us first notice that this process is
qualitatively different from the transitions $\ptp$
\cite{Li-Ster}--\cite{Sudacorrect} and $\pgg$ \cite{Radyu}. For the
latter form factors, the BLER approach yields finite answers in
the large $Q^2$ limit, which might however be affected by
significant corrections, even at large energies. The pQCD methods
are then designed to improve the accuracy of the computation, and
to investigate lower momentum transfers. On the other hand, we
have seen that $\btp$ form factors cannot be computed in the usual
factorization approach, since singular long-distance effects
arise. The pQCD approach would therefore not only improve the
precision of the BLER framework, but more importantly it would
regulate the divergences. In
refs.~\cite{pqcd}--\cite{Li-Yu} it is claimed that indeed
accurate and reliable results can be derived with no breakdown of
the perturbative framework.

The main motivation for this paper is to examine the basis for
such a strong claim. In the following sections, we  examine the
various elements involved in eq.~(\ref{eq:btpconvol}), and
emphasize our concerns about the validity and reliability of the
pQCD approach~\footnote{The pQCD approach that we have outlined is
referred to as ``$k_\perp$-resummation''. Recently, it has been
proposed to combine it with a second (``threshold'') resummation
to study $B$-decays~\cite{powercorr,Li-Sanda,Lithresh}. The concerns we
express in sections~\ref{sec:sudakovpionda} and \ref{sec:bdas}
about the pQCD approach do not depend on whether the threshold
resummation has been performed or not, and we have checked that
our numerical analysis in section~\ref{sec:numerical} is almost
unchanged when both resummations are considered. We therefore
focus on the $k_\perp$-resummation in the remainder of this
article.}.

\section{Sudakov effects in the distribution amplitude of the pion}
\label{sec:sudakovpionda}

In this section we consider the first ingredient of
eq.~(\ref{eq:btpconvol}), the distribution amplitude of the pion
$\Psi_\pi$, including Sudakov effects. The pion's  momentum is
taken to be in the $-$ direction,
$p^\prime\equiv(0,Q/\sqrt{2},\pv{0})$ (where the components are in
the $+$, $-$ and transverse directions respectively), and we
consider a configuration in which the valence quarks have
longitudinal momenta $xp^\prime$ and $\bar{x}p^\prime$ and a
transverse separation $b$. The authors of ref.~\cite{Botts-Ster}
obtain the following expression for the distribution amplitude
$\Psi_\pi$ in an axial gauge [$n\cdot A=0$, $n^2\neq 0$]:
\begin{equation} \label{eq:genbotts}
\Psi_\pi(x,b,Q;\mu)
 =\exp\left[-{\mathcal S}(n,Q,b)\right]
  \exp\left[-{\mathcal E}(\mu,b)\right]\phi_\pi(x;\mu=1/b)
    +O(\as(1/b))\ ,
\end{equation}
where:
\begin{itemize}
\item
${\mathcal S}$ is the Sudakov factor, which contains the resummed
double logarithmic contributions, as well as some non-leading
logarithms which are expected to be significant. It depends on the
transverse separation $b$, on the pion's momentum $p^\prime$ and
on the gauge-fixing vector $n$;
\item
${\mathcal E}$ is an evolution factor, relating the
renormalization scale $\mu$ and $1/b$;
\item
$\phi_\pi$ is the standard BLER distribution amplitude.
\end{itemize}

Eq.~(\ref{eq:genbotts}) exhibits two important features. The first
one is the gauge dependence of the Sudakov term ${\mathcal S}$.
Since the convolution in eq.~(\ref{eq:btpconvol}) yields physical
(gauge-independent) form factors, the gauge dependence of the
distribution amplitude $\Psi_\pi$ has to be cancelled by that in
the hard-scattering kernel. The second issue is the presence of
$O(\as)$ corrections in eq.~(\ref{eq:genbotts}). The modified
distribution amplitude $\Psi_\pi$ can only be replaced by the
standard one $\phi_\pi$ (as in eq.~(\ref{eq:genbotts})\,) for small
transverse separations $b$. Specifically, there are relative
$O(\alpha_s(1/b))$ corrections to eq.~(\ref{eq:genbotts}). Although
the Sudakov factor should suppress the full distribution amplitude
at large $b$, if this suppression is weak (or negligible), the
expression (\ref{eq:genbotts}) can not be used reliably in this
region. We investigate this numerically in
section~\ref{sec:numerical} below.

We will return to these issues later, when we consider $\btp$ form
factors. We now continue with the discussion of Sudakov effects in
the pion's distribution amplitude and summarize the analysis
performed in ref.~\cite{Botts-Ster}, and then discuss the
gauge-dependence of the Sudakov factor.

\subsection{Derivation of the Sudakov factor for the pion}
\label{subsec:derivationpi}

Sudakov effects are expected to suppress configurations with large
transverse separations. The explicit expression arises from the
resummation and subsequent exponentiation of double logarithms.
Such logarithms are related to the exchange of soft and collinear
gluons and double logarithmic contributions are due to an overlap
of both types of divergence.

Axial gauges $n\cdot A=0$ (where $n^\mu$ is a fixed vector in the
$(+,-)$ plane  and $n^2\neq 0$) are particularly suitable for use
in studies of Sudakov effects. In these gauges, the overlap arises
when a gluon is exchanged between the valence quarks of the same
meson [two-particle reducible diagrams]. Sudakov effects can
therefore be analyzed independently of the physical process, and
are included fully in the distribution amplitude as can be seen in
eq.~(\ref{eq:genbotts}). It would however be reassuring to have
an explicit demonstration of the gauge invariance of the
form-factor in which the gauge-dependence of the distribution
amplitudes is shown to cancel that of the hard scattering kernel.

In spite of the technicalities it is instructive to recall the
derivation of
eq.~(\ref{eq:genbotts})~\cite{Li-Yu,Li-Ster,Sudacorrect,Collins-Soper,
Botts-Ster}. As mentioned above, the distribution
amplitude in $x$ and impact parameter space, $\Psi_\pi$, is
defined in an axial gauge, and is gauge dependent. The only large
invariant parameter that can be constructed in the axial gauge is
$\nu^2=(p^{\,\prime} \cdot n)^2/|n^2|$. The distribution amplitude
is therefore a function $\Psi_\pi(x,b,Q;\mu)$, where the
$Q$-dependence is linked to the gauge dependence, and $\mu$ is a
renormalization scale. As we shall see below, the gauge dependence
can be used to determine the $Q$-dependence.

Eq.~(\ref{eq:genbotts}) is obtained by studying the $Q$-dependence
of $\Psi_\pi(x,b,Q;\mu)$. This distribution amplitude is expected
to exhibit the structure $\Psi_\pi(x,b,Q;\mu)=\exp[-{\mathcal
S}(x,b,Q)]$$\times$ $\Psi^0_\pi(x,b;\mu)$, where  $\Psi^0_\pi$ is
independent of $Q$, and ${\cal S}$ is found to be a RG-invariant
function containing double logarithms of $Q$. Taking such an
exponential form as an ansatz, we have
\begin{equation}
Q\frac{d\Psi_\pi}{dQ}=-Q\frac{d{\mathcal S}}{dQ} \Psi_\pi \ .
\end{equation}
${\mathcal S}$ contains at most double logarithmic terms in $Q$.
$Q\, d\mathcal{S}/dQ$ therefore contains at most single logarithms
of $Q$, which can be treated by usual renormalization-group
methods.

To obtain the pion's Sudakov factor, we have therefore to find a
differential equation of the form, $Q\,d\Psi_\pi/dQ=C\cdot
\Psi_\pi$ (where $C$ contains single logarithms of $Q$) and to
integrate this equation. The procedure can be summarized in the
following steps:
\begin{figure}[t]
\begin{center}
\includegraphics[width=11cm]{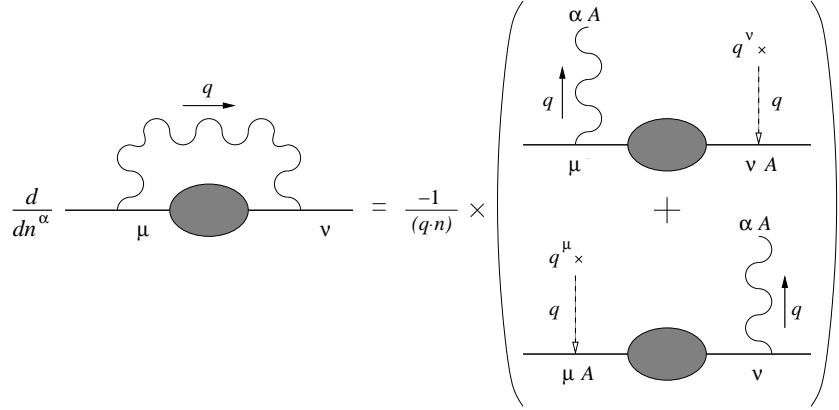}
\caption{{\footnotesize Diagrammatic representation of the
application of eq.~(\ref{eq:dndn}) for the derivative of a gluon
propagator with respect to $n^\alpha$. The dashed arrow represent
the injection of momentum $q$, using the standard vertex
$-ig\gamma_\rho T^A$ multiplied by $q^\rho$ ($\rho=\mu$ or
$\nu$).}} \label{fig:sudader}
\end{center}
\end{figure}

\begin{figure}[t]
\begin{center}
\includegraphics[width=12cm]{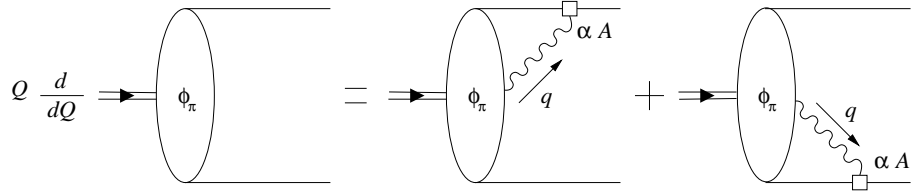}
\caption{{\footnotesize Diagrammatic representation of the
derivative of $\Psi_\pi$ with respect to $\log Q$. The square
vertex corresponds to $-igT^A \frac{n^2}{(p^\prime \cdot n)(q\cdot
n)} {p^\prime}^\alpha$.}} \label{fig:sqvertex}
\end{center}
\end{figure}

\begin{itemize}
\item $\Psi_\pi$ depends on $Q$ only through $\nu^2=(p^{\,\prime}
\cdot n)^2/|n^2|$, so that
\begin{equation} \label{eq:equivqn}
Q\frac{d\Psi_\pi}{dQ}=-\frac{n^2}{p^{\,\prime}\cdot n}
  {p^{\,\prime}}^\alpha\,\frac{d\Psi_\pi}{dn^\alpha}\ .
\end{equation}
\item The $n$-dependence of Feynman diagrams is contained only in the
gluon propagators. In the axial gauge, the propagator of a gluon
with momentum $q$ is
\begin{equation}
N^{\mu\nu}(q)=\frac{-i}{q^2}
 \left(g_{\mu\nu}-\frac{n^\mu q^\nu+q^\mu n^\nu}{n\cdot q}
   +n^2\frac{q^\mu q^\nu}{(n\cdot q)^2}\right)\ ,
\end{equation}
and the derivative with respect to $n^\alpha$ can be rewritten in
the form:
\begin{equation} \label{eq:derivgluonprop}
\frac{d}{dn_\alpha}N^{\mu\nu}(q)
  =-\frac{1}{q\cdot n}(N^{\mu\alpha} q^\nu+N^{\nu\alpha} q^\mu)\ .
\label{eq:dndn}\end{equation}
\item Consider a Feynman diagram contributing to $\Psi_\pi$.
The derivative with respect to $\log Q$ is equivalent to modifying
one of the gluon propagators according to eqs.~(\ref{eq:equivqn})
and (\ref{eq:derivgluonprop}). This is represented
diagrammatically in Fig.~\ref{fig:sudader} for the case of a gluon
propagator attached to quark lines. We need to compute the effect
of $Qd/dQ$ on all the diagrams of order $O(\as^n)$, and so we have
to consider all the diagrams of order $O(\as^{n-1})$, inserting in
all possible ways two elements: an external gluon propagator
carrying momentum $q$ (wavy line in Fig.~\ref{fig:sudader}) and an
injection of momentum $q$ (dashed arrow in
Fig.~\ref{fig:sudader}). An integral over the momentum $q$ has to
be performed.
\item When we sum over all the possible momentum injections
(represented by the dashed arrows in Fig.~\ref{fig:sudader}), Ward
identities lead to cancellations. The only surviving graphs
correspond to a momentum injection at the end of one of the
valence quark lines. Thus, applying $Qd/dQ$ means inserting a
gluon line, with one leg attached to a valence quark line with the
vertex (denoted by a square in Fig.~\ref{fig:sqvertex}):
\begin{equation}
-igT^A \frac{n^2}{(\pprime \cdot n)(q\cdot n)} {\pprime}^\alpha\ .
\end{equation}
\item For the modified diagrams, the leading contribution for large
$Q$ arises when the inserted gluon is either soft or ultraviolet.
The corresponding terms have been computed at the leading order
in $\as$ in ref.~\cite{Botts-Ster}, yielding the
differential equation:
\begin{equation}
Q\frac{d}{dQ}\Psi_\pi(x,b,Q;\mu)=
  \left[{\mathcal K}(b\mu,g(\mu))+
  \frac{1}{2}{\mathcal G}(x\nu/\mu)
   +\frac{1}{2}{\mathcal G}(\bar{x}\nu/\mu,g(\mu))\right]
   \Psi_\pi(x,b,Q;\mu)\ , \label{eq:difeqpi}
\end{equation}
where ${\mathcal K}$ and ${\mathcal G}$ collect respectively soft
and ultraviolet contributions, and have opposite anomalous
dimensions $\gamma_{\mathcal K}=-\gamma_{\mathcal G}$. The
bracketed factor in eq.~(\ref{eq:difeqpi}) is therefore
scale-independent (in ref.~\cite{Collins-Soper} it is shown that
this independence of the scale is also true at higher orders). The
explicit expressions (at order $O(\as)$) for ${\mathcal K}$ and
${\mathcal G}$ in the $\overline{\textrm{MS}}$ scheme are
\cite{Botts-Ster}:
\begin{eqnarray}
{\mathcal K}(b\mu,g(\mu))
  &=&-\frac{4}{3}\frac{\as}{\pi}\log(b^2\mu^2 e^{2\gamma}/4)\ ,
  \label{eq:oneloopk}\\
{\mathcal G}(x\nu/\mu,g(\mu))
  &=&-\frac{4}{3}\frac{\as}{\pi}
   \left[2\log\left(\frac{2x\nu}{\mu}\right)-1\right]\ ,
  \label{eq:oneloopg}
\end{eqnarray}
and~\cite{Collins-Soper,anomdimka}
\begin{equation}
\gamma_{\mathcal K}=-\gamma_{\mathcal
G}=\frac{8}{3}\frac{\as}{\pi}+
  4\left[\frac{67}{18}-\frac{\pi^2}{6}-\frac{5}{27}N_f\right]
    \left(\frac{\as}{\pi}\right)^2+O(\as^3)\ .
\end{equation}
\item The differential equation eq.~(\ref{eq:difeqpi}) can be solved
directly, since the $Q$-dependence of ${\mathcal
K}(b\mu,g(\mu))+{\mathcal G}(\bar{x}\nu/\mu,g(\mu))$ comes only
from the $\nu$-dependence of ${\mathcal G}$. A different method
was followed in refs.~\cite{Collins-Soper,Botts-Ster}. Using
RG-equations, all the $Q$-dependence can be absorbed into the
scale of the coupling constant:
\begin{eqnarray} \label{eq:rgevolkg}
{\mathcal K}(b\mu,g(\mu))+{\mathcal G}(x\nu/\mu,g(\mu))
 &=&{\mathcal K}(C_1,g(C_2x\nu))+{\mathcal G}(1/C_2,g(C_2x\nu))\\
 &&\qquad \qquad  -2\int_{C_1/b}^{C_2x\nu} \frac{d\bar\mu}{\bar\mu}
     A[C_1,g(\bar\mu)]\nono\ ,
\end{eqnarray}
where $C_1$ and $C_2$ are free parameters, which should be tuned
to avoid large logarithms in the perturbative expansion of
${\mathcal K}$ and ${\mathcal G}$ on the right hand-side of
eq.~(\ref{eq:rgevolkg}). The function $A$ is given by
\begin{equation}
A[C_1,g]=\frac{1}{2}\left[\gamma_{\mathcal K}(g)
  +\beta(g)\frac{\partial}{\partial g}{\mathcal K}(C_1,g)\right]\ .
  \label{eq:twoloopa}
\end{equation}
Combining eqs.~(\ref{eq:difeqpi}) and
(\ref{eq:rgevolkg}) yields the general solution:
\begin{eqnarray} \label{eq:solutdifeqpsi}
\Psi_\pi(x,b,Q;\mu)&=&\exp[-{\mathcal S}(x,b,\nu)]\Psi_\pi^0(x,b;\mu)
\ ,
  \\
{\mathcal S}(x,b,\nu)&=&\tilde{s}(x,b,\nu)+\tilde{s}(\bar{x},b,\nu)\ ,\\
\tilde{s}(x,b,\nu)&=&  \label{eq:defstilde}
 \frac{1}{2}\int_{C_1/b}^{C_2x\nu}
 \frac{d\bar\mu}{\bar\mu} \Bigg[
   2\log\left(\frac{C_2x\nu}{\bar\mu}\right)A(C_1,g(\bar\mu))\\
&&\qquad \qquad \qquad -{\mathcal K}(C_1,g(\bar\mu))
   -{\mathcal G}\left(\frac{1}{C_2},g(\bar\mu)\right)
   \Bigg]\ .\nonumber
\end{eqnarray}

\item
We can rewrite $\mathcal{S}$ more conveniently using
eqs.~(\ref{eq:oneloopk}), (\ref{eq:oneloopg}),
(\ref{eq:twoloopa}), and the two-loop expression for the strong
coupling constant:
\begin{eqnarray}
\frac{\as(\mu)}{\pi}&=&\frac{1}{\beta_1\log(\mu^2/\lqcd^2)}
  -\frac{\beta_2}{\beta_1^3}
      \frac{\log\log(\mu^2/\lqcd^2)}{\log^2(\mu^2/\lqcd^2)}\ ,\\
\beta_1&=&\frac{33-2N_f}{12}\ ,\qquad
\beta_2=\frac{153-19N_f}{24}\ . \label{eq:coeffbeta}
\end{eqnarray}
In this way we obtain
\begin{equation}
{\mathcal S}(x,b,\nu)=s(x,b,r\cdot Q)+s(\bar{x},b,r\cdot Q),
\end{equation}
where $r=\sqrt{|\lambda/\rho|}$ depends on the components of the
gauge-fixing vector $n$ which we write as
$n=(\lambda/\sqrt{2},\rho/\sqrt{2},\pv{0})$. The gauge dependence
now manifests itself as a dependence on $r$ rather than $\nu$.
$s(u,b,R)$ is a function of:
\begin{equation}
\hat{q}=\log\left[\frac{C_2 u R}{2\lqcd}\right],\quad \textrm{and}
 \qquad \hat{b}=\log\left[\frac{C_1}{b\lqcd}\right]\ ,
\end{equation}
and is given explicitly by:
\begin{eqnarray}
s(u,b,R)&=&
  \frac{A^{(1)}}{2\beta_1} \hat{q}
     \log\left(\frac{\hat{q}}{\hat{b}}\right)
 +\frac{A^{(2)}}{4\beta_1^2}\left(\frac{\hat{q}}{\hat{b}}-1\right)
 -\frac{A^{(1)}}{2\beta_1} (\hat{q}-\hat{b})\nono\\
&-&\frac{A^{(1)}\beta_2}{4\beta_1^3}\hat{q}
   \left[\frac{\log(2\hat{b})+1}{\hat{b}}
        -\frac{\log(2\hat{q})+1}{\hat{q}}\right]\label{eq:subr}\\
&&\hspace{-0.6in}-
 \left[\frac{A^{(2)}}{4\beta_1^2}- \frac{A^{(1)}}{4\beta_1}
    \log\left(\frac{C_1^2}{C_2^2}e^{2\gamma-1}\right)\right]
    \log\left(\frac{\hat{q}}{\hat{b}}\right)
 +\frac{A^{(1)}\beta_2}{8\beta_1^3}\left[\log^2(2\hat{q})-
 \log^2(2\hat{b})\right]
 \ .
\nono
\end{eqnarray}
The coefficients $\beta_i$ are defined in eq.~(\ref{eq:coeffbeta})
and the $A^{(i)}$'s in eq.~(\ref{eq:subr}) are given by
\begin{equation}
A^{(1)}=\frac{4}{3}\ , \qquad
A^{(2)}=\frac{67}{9}-\frac{\pi^2}{3}-\frac{10}{27}N_f
  +\frac{8}{3}\beta_1\log\left(\frac{C_1 e^\gamma}{2}\right)\ .
\end{equation}
These coefficients come from the two-loop expression of $A$ defined
in eq.~(\ref{eq:twoloopa}):
\begin{equation}
A=\frac{\as}{\pi}\,A^{(1)}+\left(\frac{\as}{\pi}\right)^2\,A^{(2)}+O(\as^3)\ ,
\end{equation}
The first three terms of $A^{(2)}$ come
from the two-loop anomalous dimension of
${\mathcal K}$~\cite{anomdimka}, and the last
one from the partial derivative of ${\mathcal K}$ with respect to $g$.

Eq.~(\ref{eq:subr}) has been derived using the two-loop level
expression of the coupling constant and of $\gamma_{\mathcal K}$,
but the one-loop expression of the functions ${\mathcal K}$ and
${\mathcal G}$. It is not necessary to compute their two-loop
expressions if $C_1$ and $C_2$ are tuned in
eq.~(\ref{eq:defstilde}) in order to avoid large contributions of
$O(\as^2)$. A conventional choice is $C_1=1$ and $C_2=\sqrt{2}$
\cite{Li-Ster}--\cite{Sudacorrect}, although we are not aware of an
argument that this choice is optimal (or even good enough for the
precision claimed for $B$-decay form factors
\cite{pqcd}--\cite{Li-Yu}).
\item Finally we need to consider the residual function
$\Psi_\pi^0(x,b;\mu)$ in eq.~(\ref{eq:solutdifeqpsi}). Since the
Sudakov factor suppresses the modified DA for large $b$, we have,
in particular, to determine $\Psi_\pi^0(x,b;\mu)$ for small $b$.
It is argued in ref.~\cite{Botts-Ster} that the factor $\exp(i\vec
k_\perp\cdot \vec b)$ in the integrand of the Fourier
transform in transverse space ensures that the integral is
dominated by the region $|k_\perp|<1/b$, and hence that
\begin{equation} \label{eq:identifpsipi}
\Psi_\pi(x,b;\mu=1/b)=\phi_\pi(x;\mu=1/b)+O(\as(1/b))\ .
\end{equation}
The logarithmic corrections of $O(\as(1/b))$ arise because the way
that the integral is regulated at large transverse momenta is
different from that in the standard definition of the BLER
distribution amplitude. A second type of correction is due to the
different definition of the BLER and modified DAs. Both quantities
are defined from the non-local, and gauge dependent matrix element
$\langle 0 | \bar{q}(z) \gamma_\mu \gamma_5 q'(0) | \pi(p)
\rangle$ in a gauge $n\cdot A=0$. But $\phi_\pi$ is defined in the
light-cone gauge [$n^2=0$], whereas $\Psi_\pi$ is defined in an
axial gauge [$n^2\neq 0$]. The identification of $\phi_\pi$ and
$\Psi_\pi$ holds only in the leading logarithmic (LL)
approximation. Thus there are next-to-leading logarithmic (NLL)
corrections in Eq.~(\ref{eq:identifpsipi}).

The presence of these corrections has important implications for
the phenomenological applications of this formalism, as will be
discussed in detail in section~\ref{sec:numerical}. In particular,
for the calculations to be valid, the integrals have to be
dominated by the region of small $b$, so that $\as(1/b)$ is in the
perturbative regime. The identification of $\Psi_\pi^0(x,b;\mu)$
with $\phi_\pi(x,\mu)$ is valid for $\mu=1/b$ and so we have to
perform the RG-evolution of $\Psi_\pi$ to the scale $\mu=1/b$. We
finally arrive at:
\begin{equation} \label{eq:rgevolpsi}
\Psi_\pi(x,b;\mu)=\exp\left[
 2\int_\mu^{1/b} \frac{d\bar\mu}{\bar\mu} \gamma_q(g(\bar\mu))
\right] \Psi_\pi(x,b;1/b)\ ,
\end{equation}
where $\gamma_q=-\as/\pi+O(\as^2)$ is the
anomalous dimension of the quark's wave function
in the axial gauges~\cite{Collins-Soper}.
\end{itemize}

Combining eqs.~(\ref{eq:solutdifeqpsi}), (\ref{eq:identifpsipi})
and (\ref{eq:rgevolpsi}), the distribution amplitude $\Psi_\pi$ is
the product of three factors: a Sudakov exponential, an
evolution-related term and the standard distribution amplitude
$\phi_\pi$. The combined expression is:
\begin{equation}
\Psi_\pi(x,b,Q;\mu)
 =e^{-{\mathcal S}}\exp\left[
 2\int_\mu^{1/b} \frac{d\bar\mu}{\bar\mu} \gamma_q(g(\bar\mu))
\right]\phi_\pi(x;\mu=1/b)+O(\as(1/b))\ . \label{eq:finalpsipi}
\end{equation}

\subsection{Dependence of the Sudakov form factors on the gauge and
on the integration constants.}

In this subsection we add some comments on the dependence of the
Sudakov factor eq.~(\ref{eq:subr}) on $C_1$, $C_2$ and $n$. The
general expression was introduced in ref.~\cite{Botts-Ster}, but
in the subsequent papers~\cite{Li-Yu,Li-Ster,Sudacorrect}, only
the particular expression with $C_1=1$, $C_2=\sqrt{2}$ and
$n\propto(1,-1,\pv{0})$ was considered.

$C_1$ and $C_2$ are arbitrary parameters, introduced in order to
solve the differential equation eq.~(\ref{eq:difeqpi}). The full
expression for $s(u,b,R)$ is therefore independent of these
parameters. However, eq.~(\ref{eq:subr}) is a truncated
expression, derived from the one-loop expression of ${\mathcal K}$
and ${\mathcal G}$ and the two-loop expression of
$\gamma_{\mathcal K}$. It is therefore weakly dependent on $C_1$
and $C_2$. A redefinition of these two coefficients does not
affect the double logarithms (i.e. the terms proportional to
$\log(\hat q/\hat b)$) but will have an effect on the single
logarithms. Ideally $C_1$ and $C_2$ should be chosen so as to
minimise the effects of the unknown higher-order terms. This
requires some insights into the important regions in $x$ and $b$.
The arguments of the logarithms in eq.~(\ref{eq:subr}) are
positive for the regions
\begin{equation}
b\leq \frac{C_1}{\lqcd} \qquad \textrm{and} \qquad
\frac{2\lqcd}{C_2 r Q} \leq x \leq
 1-\frac{2\lqcd}{C_2 r Q}\ ,
\end{equation}
and Sudakov suppression arises as $b$ reaches $C_1/\lqcd$. Let us
choose a specific gauge ($r$) and a momentum transfer ($Q$). If we
consider a soft configuration for the valence quarks of the pion,
the estimated Sudakov suppression varies with the choice of $C_1$
and $C_2$. Of course this dependence on $C_1$ and $C_2$ is due to
the fact that we have truncated the series in $\as$ to obtain
eq.~(\ref{eq:subr}).

$\Psi_\pi$ was explicitly defined in a gauge-dependent way in
eq.~(\ref{eq:modifwavefn}). As we have seen above, this gauge
dependence can be used to determine the functional dependence of
the distribution amplitude on $Q$. The gauge dependence is present
in eq.~(\ref{eq:finalpsipi}) through the Sudakov term ${\mathcal
S}$. To exhibit this gauge dependence more transparently we use
our freedom to redefine the constants of integration $C_1$ and
$C_2$, writing $C_1=C^{\,\prime}_1$ and $C_2=C^{\,\prime}_2/r$, so
that
\begin{equation}
s(x,b,r\cdot Q)=s(x,b,Q)
  +\frac{A^{(1)}}{2\beta_1}\log r
  \log\left(\frac{\log[C'_2xQ/(2\lqcd)]}
                 {\log[C'_1/(b\lqcd)]}\right)+...
\label{eq:explicit}\end{equation}
where the ellipsis denotes single logarithms. In
eq.~(\ref{eq:explicit}) we see explicitly the gauge dependence of
the Sudakov factor.

Of course the convolutions which yield physical form factors, such
as eq.~(\ref{eq:modifconvol}), are gauge independent. The gauge
dependence of the distribution amplitudes has to be compensated by
a similar term in the hard-scattering kernel. For the case of the
hard elastic scattering of two mesons, the authors of
ref.~\cite{Botts-Ster} explain clearly how the resummation of soft
gluon interactions in the hard-scattering kernel leads to a gauge
dependence cancelling the one in the meson distribution
amplitudes. We are not aware of similar explicit demonstrations
for the electromagnetic form factor of the pion or for the $\btp$
semileptonic decay form factors. Such demonstrations would add
confidence that all Sudakov effects are correctly included in the
mesons' distribution amplitudes.

\section{$B$-meson distribution amplitude(s)}\label{sec:bdas}

In section~\ref{subsec:mesonda} we have seen that there are two
distribution amplitudes for the $B$-meson at leading twist. In the
applications of the pQCD framework to
$B$-decays~\cite{pqcd}--\cite{Li-Yu}, one linear combination of
the two distribution amplitudes is neglected; specifically the two
distribution amplitudes are set equal to each other,
$\Psi^B\equiv\Psi_+^B=\Psi_-^B$. We consider the validity of
equating the two distribution amplitudes in
section~\ref{subsec:identify}. The remaining $B$-meson
distribution amplitude is analyzed along similar lines to that of
the pion. Consider a $B$-meson of momentum
$p=(m/\sqrt{2},m/\sqrt{2},\pv{0})$ containing a light quark of
momentum $l$. The modified distribution amplitude used in pQCD
analyses is:
\begin{equation} \label{eq:Bbotts}
\Psi^B(l_+,b,p\,;\,\mu)
 =\exp\left[-{\mathcal S'}(n,p,b)\right]
  \exp\left[-{\mathcal E}(\mu,b)\right]\phi^B(l_+;\mu=1/b)
    +O(\as(1/b))\ ,
\end{equation}
where:
\begin{itemize}
\item
${\mathcal S^\prime}$ is the Sudakov factor containing the
resummed double logarithmic contributions as well as selected
nonleading logarithms. It depends on the transverse separation
$b$, on the momentum of the $B$-meson $p$ and on the gauge-fixing
vector $n$.
\item
${\mathcal E}$ is an evolution factor, relating the
renormalization scale $\mu$ and $1/b$.
\item
$\phi^B\equiv\phi_+^B=\phi_-^B$ is the ``common'' distribution
amplitude of the $B$-meson.
\end{itemize}

The presence of the heavy $b$ quark significantly modifies the
analysis of Sudakov effects. The exponentiated double logarithms
of the Sudakov factor sum the overlapping contributions of soft
and collinear gluons. We work however, in the rest frame of the
$B$-meson so that the term ``collinear" is not defined until the
convolution for the form factor is introduced (with a hard
scattering kernel which is independent of $l_-$). The definition
of the distribution amplitude therefore cannot be completely
separated from the process being studied. Moreover, all the
components of the momentum of the light quark are $O(\lqcd)$, so
that the kinematics is very different from that of an energetic
pion with a light-like four-momentum. In view of these questions
we examine the derivation of eq.~(\ref{eq:Bbotts}) in
section~\ref{subsec:bderivation} below. We start however, by
considering whether setting $\Psi^B_+=\Psi^B_-$ is theoretically
motivated.

\subsection{Two distribution amplitudes or one?}
\label{subsec:identify}

In refs.~\cite{pqcd}--\cite{Li-Yu} the phenomenological analysis
is performed with a single distribution amplitude for the B-meson,
setting $\phi^B\equiv\phi_+^B=\phi_-^B$ and
$\Psi^B\equiv\Psi_+^B=\Psi_-^B$. Such an identification simplifies
considerably the calculation of the form factors and other
physical quantities, since the momentum-space projection operator
$P^B$ in eq.~(\ref{eq:bmesonproj}) now reduces to:
\begin{equation}
P^B_{\beta\alpha}=-\frac{if_B}{4}
  \left[(\dirac{p}+m)\gamma_5\right]_{\alpha\beta}
  \phi^B(\omega)\ .
\end{equation}
In view of the arguments presented in sec.~\ref{subsec:mesonda}
above, we find this identification of the two distribution
amplitudes rather surprising. Indeed the authors of
ref.~\cite{Groz-Neub} argue that the two DAs are likely to have a
different end-point behaviour. It would be surprising to us if the
pseudoscalar matrix element in eq.~(\ref{eq:pseudomatelem}) and
the axial one in eq.~(\ref{eq:axialmatelem}) could be described in
the heavy-quark limit by the same distribution amplitude. In this
section we examine the weaker proposition in ref.~\cite{Li-Sanda};
that although $\phi_+^B$ and $\phi_-^B$ are different, they have
the same contribution to the convolution (\ref{eq:btpconvol}), so
that that their difference can be neglected when studying $\btp$
form factors.

In order to assess this claim, we need the $\btp$ form
factors~(\ref{eq:f0std}) and (\ref{eq:fpstd}) written as
convolutions over both longitudinal and transverse momenta:
\begin{equation}
F_{0,+} = f_\pi f_B m^2 \frac{\as\pi C_F}{N_c}
 \int_0^1 dx \int dl_+ \int d^2\pv{k} \int d^2\pv{l}
    {\mathcal I}_{0,+}\ ,
\label{eq:f0p}\end{equation} where
\begin{eqnarray}
{\mathcal I}_{0}&=&
  \frac{1}{\bar{x}l_+\eta m+(\pv{k}+\pv{l})^2} \eta\Psi_\pi
  \nonumber\\
&\times &\Bigg\{
  \frac{1}{\bar{x}\eta m^2+\pv{k}^2}
  \left[\Psi_-^B+\bar{x}\eta \Psi_+^B
    -\frac{\Delta(l_+,\pv{l})}{m} \frac{\pv{k}\cdot (\pv{k}+\pv{l})}
      {\bar{x}l_+\eta m+(\pv{k}+\pv{l})^2}\right]\label{eq:eqi0}\\
&&\hspace{-0.3in}+
   \frac{1}{l_+\eta m+\pv{l}^2}
 \left(\bar\eta \frac{l_+}{m}\Psi_+^B+\frac{\Delta(l_+,\pv{l})}{m}
 \left[1-\frac{\pv{l}^2}{l_+\eta m+\pv{l}^2}
  -\frac{(\pv{k}+\pv{l})\cdot\pv{l}}
      {\bar{x}l_+\eta m+(\pv{k}+\pv{l})^2}\right]\right)
  \Bigg\}\ ,\nono
\end{eqnarray}
and
\begin{eqnarray}
 {\mathcal I}_{+}&=& \label{eq:eqip}
  \frac{1}{\bar{x}l_+\eta m+(\pv{k}+\pv{l})^2} \Psi_\pi
  \nonumber\\
&\times& \Bigg\{
  \frac{1}{\bar{x}\eta m^2+\pv{k}^2}
  \left[\Psi_-^B+\bar{x}\eta \Psi_+^B
    -\frac{\Delta(l_+,\pv{l})}{m} \frac{\pv{k}\cdot (\pv{k}+\pv{l})}
      {\bar{x}l_+\eta m+(\pv{k}+\pv{l})^2}\right]\\
&&\hspace{-0.3in} +
   \frac{1}{l_+\eta m+\pv{l}^2}
 \left(-\bar\eta \frac{l_+}{m}\Psi_+^B+\frac{\Delta(l_+,\pv{l})}{m}
 \left[1-\frac{\pv{l}^2}{l_+\eta m+\pv{l}^2}
  -\frac{(\pv{k}+\pv{l})\cdot\pv{l}}
      {\bar{x}l_+\eta m+(\pv{k}+\pv{l})^2}\right]\right)
  \Bigg\}\ ,\nono
\end{eqnarray}
with $\Delta(\omega,\pv{l})=\int_0^\omega d\ell
  [\Psi_-^B(\ell,\pv{l})-\Psi_+^B(\ell,\pv{l})]$.
$\Psi_\pm^B$ are functions of $l_+$ and $\pv{l}$, and $\Psi_\pi$
of $x$ and $\pv{k}$. In the formula for $\mathcal{I}_0$, the
contribution in the second line comes from the upper diagram of
Fig.~\ref{btopilowest}, and the third line from the lower diagram.
The second term in the formula for $\mathcal{I}_+$ comes from the
lower diagram.

The authors of ref.~\cite{Li-Sanda} suggest that the value of
$F_{0,+}$ should remain stable if we replace $\Psi_+^B$ by
$\Psi_-^B$ (or vice-versa) in eqs.~(\ref{eq:eqi0}) and
(\ref{eq:eqip}). To support this idea, they introduce the
following ``reasonable parametrization" for the distribution
amplitudes:
\begin{equation}
\phi_+^B(l_+)=\delta(l_+-\Lambda)-\Lambda\delta'(l_+-\Lambda)\
,\qquad \qquad \phi_-^B(l_+)=\delta(l_+-\Lambda)\ ,
\label{eq:phibdelta}
\end{equation}
where $\Lambda$ is a parameter of order $\lqcd$.
After integrating $\mathcal{I}_0$ and $\mathcal{I}_+$ over the
transverse momentum $\pv{l}$, the hard-scattering
kernels behave like $\ln (l_+/m)$ for small $l_+$. Convoluting
them with $\phi_-^B$ and $\phi_+^B$, we obtain two equal
contributions  proportional to $\ln(m/\Lambda)$, but with
different constant terms. It is therefore argued that the two $B$
meson DAs can be set equal to each other in $\btp$ form factors
with reasonable precision.

The above argument however, raises a number of important
questions. First of all we note that the two $B$-meson DAs in
eq.~(\ref{eq:phibdelta}) do not satisfy the equation of motion
eq.~(\ref{eq:eom}). We also note that this argument implicitly
assumes that the dependence on the transverse momentum of the
modified DAs $\Psi^B_\pm$ is approximately flat. Of course we
accept that the model in eq.~(\ref{eq:phibdelta}) was introduced
only for purposes of illustration, nevertheless we do not consider
it to be sufficiently general to support the case that we can
identify $\phi_+^B$ and $\phi_-^B$, and use a single DA in the
analysis. To underline this point we introduce an alternative (and
equally unrealistic) model to that in eq.~(\ref{eq:phibdelta}),
but one which satisfies the equation of motion (\ref{eq:eom}):
\begin{equation}\label{eq:phibtheta}
\phi_+^B(l_+)=\delta(l_+-\Lambda) \qquad\textrm{and} \qquad
\phi_-^B(l_+)=\frac{1}{\Lambda}\Theta(l_+)\ ,
\label{eq:alternative}\end{equation}
where $\Lambda=O(\lqcd)$, and we define $\Theta(l_+)=1$ when
$0\leq l_+\leq \Lambda$, and zero outside this range. We now
perform the following simple exercise. We take the first term in
eq.~(\ref{eq:eqi0}) (proportional to $\Psi_-^B$) and assume (as
above) that its dependence on the transverse momentum is
approximately flat over $[0,\Lambda]$. We then study the
contributions to ${\cal I}_0$ with $\phi_-^B$ given by each of the
two functions in eq.~(\ref{eq:phibtheta}).  In this way we
investigate whether it matters which of $\phi_-^B$ or $\phi_+^B$
is used in the calculation (at least for this one term).

If we use $\phi_-^B(l_+)=\Theta(l_+)/\Lambda$, we can perform the
integration  over $l_+$ and $\pv{l}$:
\begin{equation}
f_\pi f_B m^2 \frac{\alpha_s \pi C_F}{N_C}
 \int_0^1 dx \int d^2 \pv{k}
 \Psi_\pi \frac{1}{\bar{x}m^2+\pv{k}^2} c_-(\bar{x},k_\perp)\ ,
 \label{eq:contribtheta}
\end{equation}
where $c_-$ is a function of $\bar{x}$ and $K=k_\perp/\Lambda$:
\begin{eqnarray}
c_-&=&\frac{1}{2\Lambda^2\bar{x}}\Bigg[L|1-K^2|- S
  -2L \log[1+K^2+|1-K^2|]\\
&&\quad  +\bar{x} - 2\bar{x}\log[2\bar{x}]
  +2\bar{x} \log[\bar{x}+S+L(1-K^2)]
  +2L\log[\bar{x}+S+L(1+K^2)]\Bigg]\ ,\nono
\end{eqnarray}
and $L=\Lambda/m$ and $S=\sqrt{[\bar{x}+L^2(1+K^2)]^2-4K^2L^2}$.

If we use $\phi_-^B(l_+)=\delta(l_+-\Lambda)$, we obtain a similar
expression, where $c_-$ is replaced by:
\begin{equation}
c_+=\frac{1}{\Lambda^2}
  \left[\log[\bar{x}+S+L(1-K^2)]-\log[2\bar{x}]\right]\ .
\end{equation}
From eq.~(\ref{eq:contribtheta}) we see that $c_-$ and $c_+$ have
to be convoluted with $\Psi_\pi/(\bar{x}m^2+\pv{k}^2)$ in order to
compute the contribution of $\Psi_-^B$ to the form factor
$F_0(0)$. When the integration over $x$ and $\pv{k}$ is performed,
the largest contribution comes from the region of small $\bar{x}$
and $k_\perp$.

For small $L=\Lambda/m$, the behaviour of $c_-$ and $C_+$ is given
by:
\begin{equation} \label{eq:smallL}
c_-\sim -\frac{\log L}{m^2 L \bar{x}}\ ,
\qquad
c_+\sim \frac{1}{m^2 L \bar{x}}
\end{equation}
which will be convoluted with the pion DA in
eq.~(\ref{eq:contribtheta}). We see therefore that the model in
eq.~(\ref{eq:phibdelta}) was artificially designed in order to
yield the same leading contribution $\ln(\lqcd/m)$. In general,
the contributions from $\phi_+^B$ and $\phi_-^B$ have different
leading logarithmic behaviour and are significantly different in
size. In the model defined by eq.~(\ref{eq:alternative}), the
expressions in eq.~(\ref{eq:smallL}) suggest that $\phi_-^B$
yields a much larger contribution than $c_+$ and we now check this
numerically.

\begin{figure}[t]
\begin{center}
\includegraphics[width=11cm]{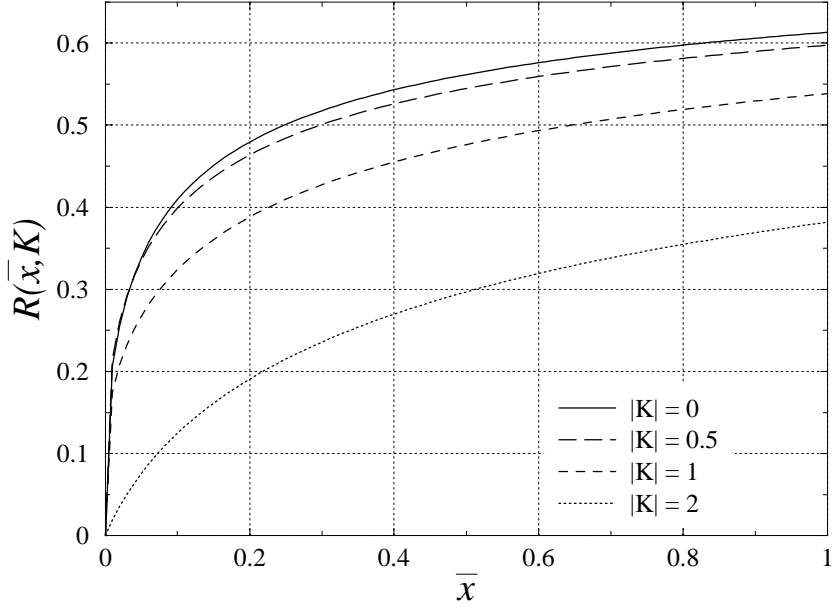}
\caption{{\footnotesize Ratio $R=\frac{c_--c_+}{c_-+c_+}$, as a function
of $\bar{x}$, for different values of $K=k_\perp/\Lambda$
$[\Lambda/m=0.25/5.28]$.}}
\label{fig:err}
\end{center}
\end{figure}

We introduce the ratio $R=(c_--c_+)/(c_-+c_+)$ which depends on
$\bar{x}$ and $K$. The identification $\phi^B=\phi_+^B=\phi_-^B$
is justified if $R$ is small in the region which gives the
dominant contribution to the convolution
eq.~(\ref{eq:contribtheta}). In fig.~\ref{fig:err} we plot the
ratio $R$ as a function of $\bar{x}$ for different values of
$K=k_\perp/\Lambda$. For purposes of illustration we take
$L=\Lambda/m=0.25/5.28$. We see that $R$ can be as large as 40~\%
for $k_\perp\leq \Lambda$, even for very small $\bar{x}$. The
identification of the two $B$-meson DAs would only be valid for
$\bar{x}<0.05$, or $k_\perp \gg \Lambda$, i.e. in regions which do
not contribute very much to the convolution
eq.~(\ref{eq:contribtheta}).

The analysis of a different, and perhaps slightly more realistic
model, introduced in Appendix~\ref{sec:errest}, confirms the
possibility of a large error. Of course, we have not estimated the
actual error on $F_{0,+}$ when $\Psi^B_+$ and $\Psi^B_-$ are set
equal to each other. In this presentation we have considered only
a part of the contribution to the form factor $F_0(0)$, the scale
of the coupling constant has not been defined, and Sudakov effects
have not been included. Nevertheless, lessons can be learned from
this brief analysis. The hard-scattering kernel enhances the small
$l_+$ region, where the $B$-meson DAs are concentrated. If
$\phi_+^B$ and $\phi_-^B$ have different end-point behaviours, the
difference may be strongly enhanced by the hard-scattering kernel.
There is therefore no convincing motivation for setting the two
DAs equal to each other and such a procedure is likely to lead to
unreliable results.

A related issue is the contribution from
$\Delta(\omega)=\int_0^\omega d\ell \,(\phi^B_-(\ell) -
\phi^B_+(\ell)\,)$. The results of the standard approach,
eqs.~(\ref{eq:f0std}) and (\ref{eq:fpstd}), can be obtained by
neglecting the transverse momenta in eqs.~(\ref{eq:eqi0}) and
(\ref{eq:eqip}). We see therefore that a term proportional to
$\Delta$ is contained in eqs.~(\ref{eq:f0std}) and
(\ref{eq:fpstd})~\footnote{This contribution is however not
explicit in eqs.~(\ref{eq:f0std}) and (\ref{eq:fpstd}) because we
have used the equation of motion eq.~(\ref{eq:eom}) to write
$\Delta(l_+)=l_+\phi_-^B(l_+)$.}. The contribution of $\Delta$ to
the standard formulae eqs.~(\ref{eq:f0std}) and (\ref{eq:fpstd})
is divergent when $l_+\to 0$ or $\bar{x}\to 0$ and a large
contribution from this region is therefore also expected in the
pQCD approach. This contribution is dropped however, if one sets
$\Psi_+^B= \Psi_-^B$ (and hence $\Delta=0$).

We conclude that the model approximation in which the two
distribution amplitudes are set equal to each other is not a
generic one and in general will lead to wrong results for physical
quantities.

\subsection{Derivation of the Sudakov factor for the $B$-meson}
\label{subsec:bderivation}

In spite of the strong reservations expressed in the previous
subsection, let us suppose that we can consider a single $B$-meson
distribution amplitude. In this subsection we then examine the
derivation of the Sudakov factor for the DA of the $B$-meson given
in ref.~\cite{Li-Yu} and used in subsequent phenomenological
analyses. According to the authors of ref.~\cite{Li-Yu}, there can
be an overlap of soft and collinear divergences in gluon
exchanges, when $l_+$ is much larger than $\lqcd$, i.e. of order
$M_B$. These divergences can be resummed into a Sudakov factor,
which is therefore  ``half'' of the Sudakov factor for the pion,
since there is one light valence quark instead of two. This factor
would then lead to a Sudakov suppression of configurations with
large transverse separations.

\begin{figure}[t]
\begin{center}
\includegraphics[width=13cm]{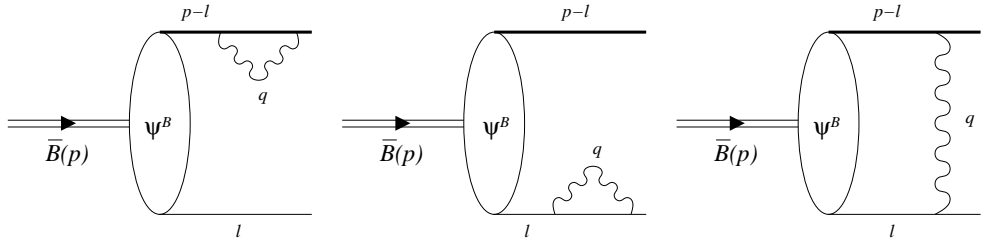}
\caption{{\footnotesize $O(\as)$ corrections to the distribution
amplitude for the $B$-meson.}}
\label{fig:corrphiB}
\end{center}
\end{figure}

We work in the rest frame of the $B$-meson in which the light
valence quark has components of momentum of the same order in all
directions [$O(\lqcd)$]. The term ``collinear'' therefore is not
meaningful at this stage and no double logarithms are expected to
arise, as noted by the authors of ref.~\cite{kroll}. If one of the
light-quark components, $l_+$ say, is much larger than $\lqcd$,
the possible Sudakov suppression happens for configurations that
already highly suppressed. The impact of Sudakov effects is
therefore likely to be rather weak. On the other hand, in the most
likely configuration where all the light-quark components are
small [$O(\lqcd)$], no Sudakov suppression is expected. There is
therefore no perturbative QCD mechanism to suppress configurations
with a large transverse separation but a small longitudinal
momentum. The consequence of this fact will be investigated in
section~\ref{sec:numerical}.

The argument in ref.~\cite{Li-Yu} attempts to proceed along the
same lines as were outlined for the pion in
section~\ref{subsec:derivationpi}, by studying the dependence of
the $B$-meson DA on $l_+$. We now briefly note, without comment,
some of the relevant features of this argument, stressing in
particular the differences with the corresponding derivation for
the pion. We then present our criticisms of this derivation.
\begin{itemize}
\item The authors of ref.~\cite{Li-Yu} aim to identify the
invariant parameters on which the distribution amplitude
$\Psi^B=\Psi_+^B=\Psi_-^B$ can depend. Because of the presence of
the heavy quark, many such invariant parameters are available:
$p^2$, $p\cdot l$, $p \cdot n$, $l \cdot n$ and $n^2$. It is still
true that the structure of the gluon propagator in the axial gauge
leads to a scale invariance of $\Psi^B$ with respect to $n$, but
this argument is not sufficient to deduce that $\Psi^B$ depends on
a single parameter.
\item It is then argued that the eikonal approximation can be used to
overcome the difficulties caused by the presence of the heavy
quark. Consider a gluon of momentum $q$ attached to the heavy
quark line of momentum $p-l$. When $q$ is soft, or collinear to
$l$, we can use the approximation:
\begin{equation}
\frac{\dirac{p}-\dirac{l}+\dirac{q}+m}{(p-l+q)^2-m^2}\gamma_\alpha
  =\frac{p^\alpha}{p\cdot q}+R\ . \label{eq:eikon}
\end{equation}
$R$ contains terms that are suppressed by powers of $1/m$ or that
vanish once contracted with $(\dirac{p}+m)$.
\item
The eikonal approximation leads to a scale invariance of $\Psi^B$
with respect to $p$. $p$ would only appear in ratios such as
$(n\cdot p)^2/(p^2n^2)$, which are disregarded because they cannot
lead to a ``large scale''. The only ``large scale'' would be $l$,
and the relevant invariants would be $(l\cdot n)^2/n^2$ and
$(l\cdot p)^2/p^2$.

The $O(\as)$ corrections are examined in the particular axial gauge
$n=(1,-1,\pv{0})$,
in which $\Psi^B$ (Fig.~\ref{fig:corrphiB}) is found to exhibit a
scale invariance with respect to $l$ in the same eikonal
approximation. The authors of ref.~\cite{Li-Yu} conclude that
$(l\cdot p)^2/p^2$ has to be discarded and $\Psi^B$ depends only
on the scale ${\nu\,^\prime}^2=(l\cdot n)^2/n^2$.
\item Since $\Psi^B$ depends on a single scale $\nu\,^\prime$,
it is possible to replace $l_+\,d/dl_+$ by $d/dn^\alpha$, and
to derive a differential equation for the $B$-meson distribution
amplitude:
\begin{equation}
l_+\frac{d}{dl_+}\Psi^B(l_+,b;\mu)=
  \frac{1}{2}\left[{\mathcal K}(b\mu,g(\mu))+
  {\mathcal G}(\nu\,^\prime/\mu)\right]
   \Psi^B(l_+,b;\mu)\ . \label{eq:difeqB}
\end{equation}
\item
This equation can be solved using the same techniques used for the
pion in section~\ref{subsec:derivationpi} leading to:
\begin{equation}
\Psi^B(l_+,b\,;\mu)
 =e^{-\tilde{s}(1,b,\nu\,^\prime)}\exp\left[
 2\int_\mu^{1/b} \frac{d\bar\mu}{\bar\mu} \gamma_q(g(\bar\mu))
\right]\phi^B(l_+;\mu=1/b)+O(\as(1/b))\ ,
\end{equation}
where $\tilde{s}$ is defined in eq.~(\ref{eq:defstilde}). We can
rewrite the Sudakov factor in terms of $l_+$:
$\tilde{s}(1,b,\nu\,^\prime)=s(l_+/m,b,m/r)$, where $s$ is defined
in eq.~(\ref{eq:subr}).
\end{itemize}

We now express our concerns about this derivation, which include
the identification of the two $B$-meson distribution amplitudes,
the use of an eikonal approximation for the heavy quark and the
determination of the scales on which the $B$-meson distribution
amplitude depends.
\begin{itemize}
\item It is not clear to us that multiplicative, RG-invariant
Sudakov factors arise separately for
$\Psi^B=(\Psi_+^B+\Psi_-^B)/2$ and
$\bar\Psi^B=(\Psi_+^B-\Psi_-^B)/2$ (the latter is neglected in
ref.~\cite{Li-Yu}).
\item
The eikonal approximation eq.~(\ref{eq:eikon}) is relevant when
$l_+$ is soft. But the resummation of double logarithms is
performed in the region where $l_+$ is large. The relevance of
eq.~(\ref{eq:eikon}) in this case is not clear to us.
\item A demonstration of the $l$-scale invariance of the $O(\as)$
corrections to $\Psi^B$, in a particular gauge, does not ensure
the scale invariance of the function $\Psi^B$. And even if this
property was satisfied by $\Psi^B$, $(l\cdot p)^2/p^2$ and
$(l\cdot n)^2/n^2$ are both $l$-scale dependent and therefore
could not be relevant variables for $\Psi^B$. If $\Psi^B$ is scale
independent with respect to $n$, $p$ and $l$, it cannot depend on any
``large scale''. The invariants with this property are
\begin{equation}
(p\cdot n)^2/[p^2 n^2]=\frac{\rho}{4\lambda}
  \left(1+\frac{\lambda}{\rho}\right)^2\ ,
\qquad (l\cdot p)^2n^2/[p^2(l\cdot n)^2]=\frac{\lambda}{\rho}\ ,
\end{equation}
where $n=(\lambda/\sqrt{2},\rho/\sqrt{2},\pv{0})$ and only the
$l_+$ component of $l$ has been retained.
\end{itemize}

Moreover, the questions raised by the derivation of the pion
Sudakov factor remain to be answered:
\begin{itemize}
\item
How is the gauge dependence of $\Psi^B$ cancelled by that in the
hard-scattering kernel ?
\item Is it legitimate and sufficiently precise to identify $\phi_B$
and $\Psi^B$ ?
\end{itemize}

We have tried and failed to understand the derivation of the
Sudakov factor for the distribution amplitudes of the $B$-meson.
In our view it is based on a number of unjustified assumptions and
steps. Given that the dominant contributions come from the region
of small $l_+$, it seems to us that Sudakov effects are unlikely
to be significant for the $B$-meson. If this is the case however,
the integrals in eqs.~(\ref{eq:f0std}) and (\ref{eq:fpstd}) may
give significant contributions (or even diverge) at small $l_+$.

\section{Numerical study}
\label{sec:numerical}

In the previous sections we have explained our reservations about
the pQCD formalism used to evaluate the semileptonic form factors
and other physical quantities. We now temporarily set aside these
reservations and perform a numerical study of the pQCD evaluation
of the $\btp$ semileptonic form factors. We are particularly
interested in two questions:
\begin{enumerate}
\item Are the results sensitive to the choice of distribution
amplitudes?\\ We will see that the answer to the first question is
\textit{yes}, so that the form factors are not calculable in
practice.
\item For ``reasonable" choices of distribution amplitudes, does
all of the contribution come from the perturbative region?\\ The
answer to this question is in general \textit{no}, so that the
form factors are not calculable also in principle. This conclusion
is in addition to our theoretical reservations.
\end{enumerate}

\subsection{pQCD expression of $\btp$ form factors}

The formulae for the form factors in eqs.~(\ref{eq:f0p}),
(\ref{eq:eqi0}) and (\ref{eq:eqip}) can be rewritten in impact
parameter space:
\begin{eqnarray}
F_{0,+}&=&\int_0^1 dx \int dl_+ \int b_\perp db_\perp
  \int c_\perp dc_\perp\\
&&\qquad \qquad
  \tilde\Psi_\pi(x,b_\perp;\mu)
  \tilde{H}_{0,+}(x,l_+,b_\perp,c_\perp,\eta,m;\mu)
  \tilde\Psi^B(l_+,c_\perp;\mu)\ ,
\end{eqnarray}
where
\begin{eqnarray}
\tilde{H}_{0}&=&f_\pi f_B m^2 \frac{\pi C_F}{N_c}\nono
\alpha_s(\mu)\eta\\
&&\quad \times\Bigg[(1+\bar{x}\eta)
   G(\bar{x}l_+\eta m,b_\perp,\bar{x}\eta m^2,c_\perp)
   +\frac{l_+\bar\eta}{m}
   G(l_+\eta m,c_\perp,\bar{x}l_+\eta m,b_\perp)\Bigg]
  \  ,\nonumber
\end{eqnarray}
and
\begin{eqnarray}
\tilde{H}_{+}&=&f_\pi f_B m^2 \frac{\pi C_F}{N_c}\alpha_s(\mu)\\
&&\times \Bigg[(1+\bar{x}\eta)
    G(\bar{x}l_+\eta m,b_\perp,\bar{x}\eta m^2,c_\perp)
 -\frac{l_+\bar\eta}{m}
    G(l_+\eta m,c_\perp,\bar{x}l_+\eta m,b_\perp)\Bigg]\ ,\nono
\end{eqnarray}
with the function $G$ given in terms of Bessel functions by:
\begin{eqnarray}
G(A,b_\perp,B,c_\perp)
   &=&K_0(\sqrt{A}\,b_\perp)
     [\theta(b_\perp-c_\perp)K_0(\sqrt{B}\,b_\perp)
         I_0(\sqrt{B}\,c_\perp)\\
&&\qquad\qquad\qquad
     +\theta(c_\perp-b_\perp)K_0(\sqrt{B}c_\perp)
         I_0(\sqrt{B}\,b_\perp)]\ .
  \nono
\end{eqnarray}
The angular integrations in the transverse plane have been
performed, since the distribution amplitudes are independent of
the orientation of the transverse momenta.

There are higher order corrections to the hard-scattering
amplitude of order $\log(t/\mu)$, where $t$ is the large scale
associated with the hard gluon. In ref.~\cite{Li-Yu} this was
taken as $t=\max(\sqrt{l_+x\eta m},1/b_\perp,1/c_\perp)$. In order
to avoid large corrections, we can perform the evolution of the
hard-scattering kernels $H_{0,+}$ from $\mu$ to $t$ using the RG
equation:
\begin{equation}
\left(\mu\frac{\partial}{\partial\mu}+\beta(g)\frac{\partial}{\partial g}
  \right)H_{0,+}=4\gamma_q H_{0,+}\ ,
\end{equation}
where $\gamma_q=-\alpha_s/\pi$ is the anomalous dimension of the
quark's wave function in the axial gauge. For $H_{0,+}(t)$ we take
its expression at lowest order in $\as$. At this order $H$ does
not contain a gauge dependence able to cancel the gauge dependence
of $\Psi_\pi$ and $\Psi^B$. In particular, there is no resummation
of soft gluon contributions in the hard-scattering kernel in a
similar way to the work of ref.~\cite{Botts-Ster} for hard elastic
scattering of mesons. We obtain:
\begin{eqnarray}
F_0&=&f_\pi f_B m^2 \frac{\pi C_F}{N_c}
 \int_0^1 dx \int dl_+ \int b_\perp db_\perp
  \int c_\perp dc_\perp  \nono\\
&&\qquad \times\phi_\pi(x;1/b_\perp)
  \phi^B(l_+;1/c_\perp)\exp[-S(x,l_+,\eta,m,b_\perp,c_\perp)]
  \alpha_s(t)
  \label{eq:f0withsudakov}\\
&&\hspace{-0.3in}\times\eta\Bigg[(1+\bar{x}\eta)
   G(\bar{x}l_+\eta m,b_\perp,\bar{x}\eta m^2,c_\perp)
  +\frac{l_+\bar\eta}{m}
    G(l_+\eta m,c_\perp,\bar{x}l_+\eta m,b_\perp)\Bigg]
  \ ,\nonumber
\end{eqnarray}
and
\begin{eqnarray}
 F_+&=&f_\pi f_B m^2 \frac{\pi C_F}{N_c}
 \int_0^1 dx \int dl_+ \int b_\perp db_\perp
  \int c_\perp dc_\perp\nono\\
&&\qquad \times \phi_\pi(x;1/b_\perp)
  \phi^B(l_+;1/c_\perp)\exp[-S(x,l_+,\eta,m,b_\perp,c_\perp)]
  \alpha_s(t)
  \label{eq:fpluswithsudakov}\\
&&\hspace{-0.3in}\times \Bigg[(1+\bar{x}\eta)
    G(\bar{x}l_+\eta m,b_\perp,\bar{x}\eta m^2,c_\perp)
 -\frac{l_+\bar\eta}{m}
    G(l_+\eta m,c_\perp,\bar{x}l_+\eta m,b_\perp)\Bigg]\ ,\nono
\end{eqnarray}
where the function $S$ combines Sudakov factors and evolution-related
terms:
\begin{eqnarray}
S&=&s(x,b_\perp,\eta m r)+s(\bar{x},b_\perp,\eta m r)
  +s(l_+/m,c_\perp,m r)\\
&&\quad  +2\int_{1/b_\perp}^t \frac{d\bar\mu}{\bar\mu}
    \gamma_q(g(\bar\mu))
  +2\int_{1/c_\perp}^t \frac{d\bar\mu}{\bar\mu}
    \gamma_q(g(\bar\mu))\ .\nono
\end{eqnarray}
The function $s$ is defined in eq.~(\ref{eq:subr}). We follow
refs.~\cite{Li-Yu,Li-Ster} and take the particular values
$C_1=1$, $C_2=\sqrt{2}$ and $r=1$, so that
\begin{equation}
\hat{q}=\log[uQ/(\sqrt{2}\lqcd)],\quad \textrm{and}\quad
\hat{b}=\log[1/(b\lqcd)]\ .
\end{equation}

In eqs.~(\ref{eq:f0withsudakov}) and (\ref{eq:fpluswithsudakov}),
the transverse separations are integrated from 0 to $1/\lqcd$,
because the Sudakov factors suppress any configuration with larger
transverse separations. We can accept this statement in the case
of the pion ($b_\perp$). But the Sudakov factor for the $B$-meson
is weak for the usual configurations where $l_+$ is of order
$\lqcd$ or less. We may therefore worry that the integration with
respect to $c_\perp$ is artificially cut-off, and omits regions
which might yield significant contributions to the form factors
(such as small $l_+$ but large $c_\perp$).

\subsection{Models of distribution amplitudes}
\label{subsec:modeldas}

In order to compute the $\btp$ form factors, we have to choose
models for the distribution amplitudes of the pion and the $B$
meson. It can be seen from eq.~(\ref{eq:finalpsipi}) that the
modified pion DA $\Psi_\pi(x,b;\mu)$ is the product of three
factors: a Sudakov exponential, an evolution-related term and the
BLER pion DA $\phi_\pi(x;\mu=1/b)$. In
ref.~\cite{Li-Yu}, the $b$-dependence due to the renormalization
scale of $\phi_\pi(x;\mu=1/b)$ has been neglected in
eqs.~(\ref{eq:f0withsudakov}) and (\ref{eq:fpluswithsudakov}),
because its numerical impact is believed to be small. We will
follow this prescription here. In a similar way, the
$b$-dependence of the $B$-meson DA in
eqs.~(\ref{eq:f0withsudakov}) and (\ref{eq:fpluswithsudakov}) is
neglected.

We use two different models for the leading-twist pion
DA~\cite{Brod-Lep,Chern}:
\begin{eqnarray}
\phi_\pi^{(a)}(x) &=&
  6x(1-x)[1+\alpha_2 C_2^{(3/2)}(2x-1)+\alpha_4 C_4^{(3/2)}(2x-1)]\ ,
  \label{eq:pidawithmoments}\\
\phi_\pi^{(b)}(x) &=& 30x(1-x)(1-2x)^2\ ,\label{eq:pidacz}
\end{eqnarray}
where $C_n^{(3/2)}(u)$ are the Gegenbauer polynomials.
Even though we neglect evolution effects in the computation of the
$\btp$ form factors, the BLER distribution amplitudes do depend on
the renormalization scale $\mu$. In model $\phi_\pi^{(a)}$ the
$\mu$-dependence is contained in the coefficients $\alpha_2$ and
$\alpha_4$; model $\phi_\pi^{(b)}$ was proposed originally at the
scale $\mu_0=$0.5 GeV~\cite{Chern}.

\begin{figure}[t]
\begin{center}
\includegraphics[width=7cm]{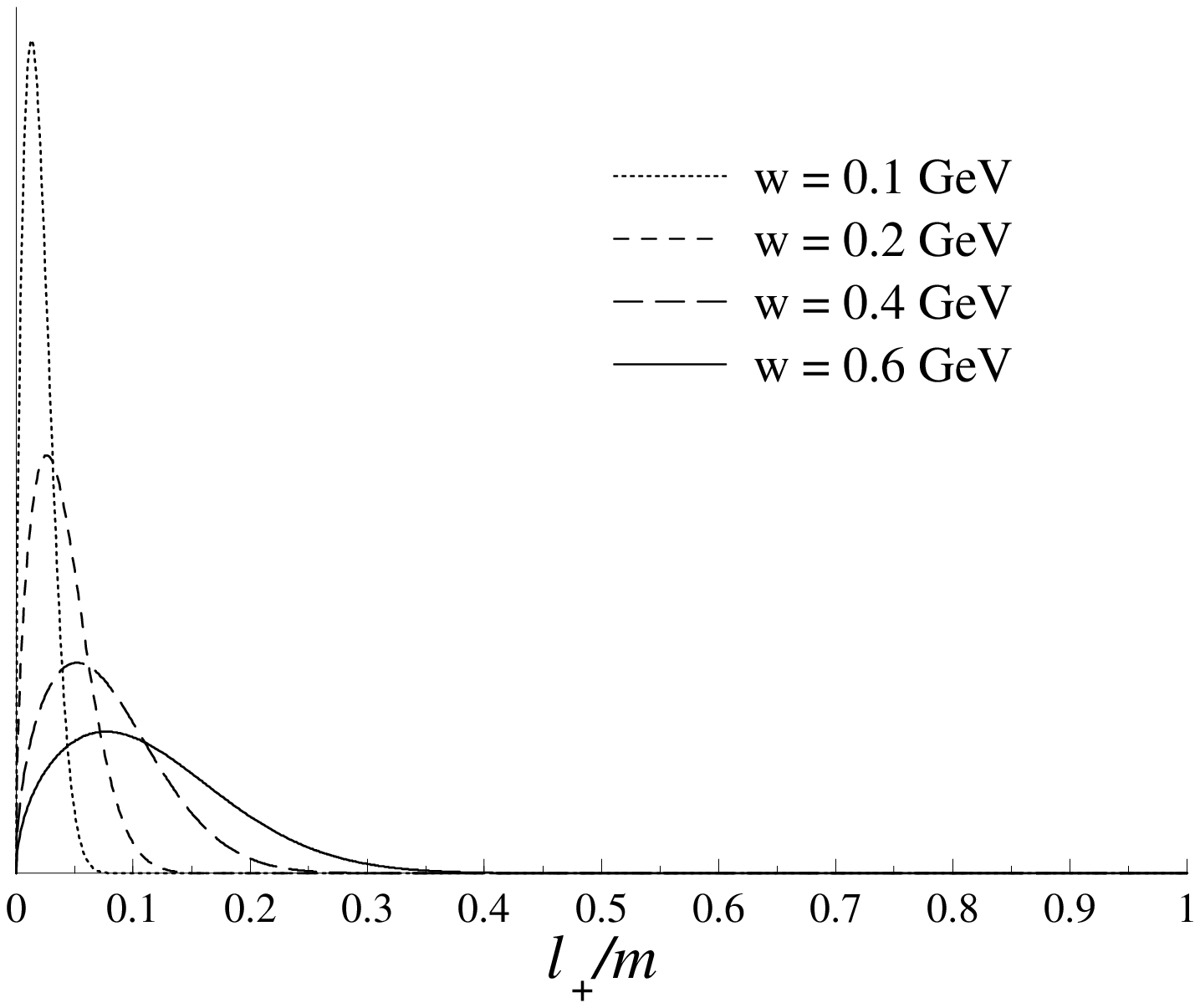}
\includegraphics[width=7cm]{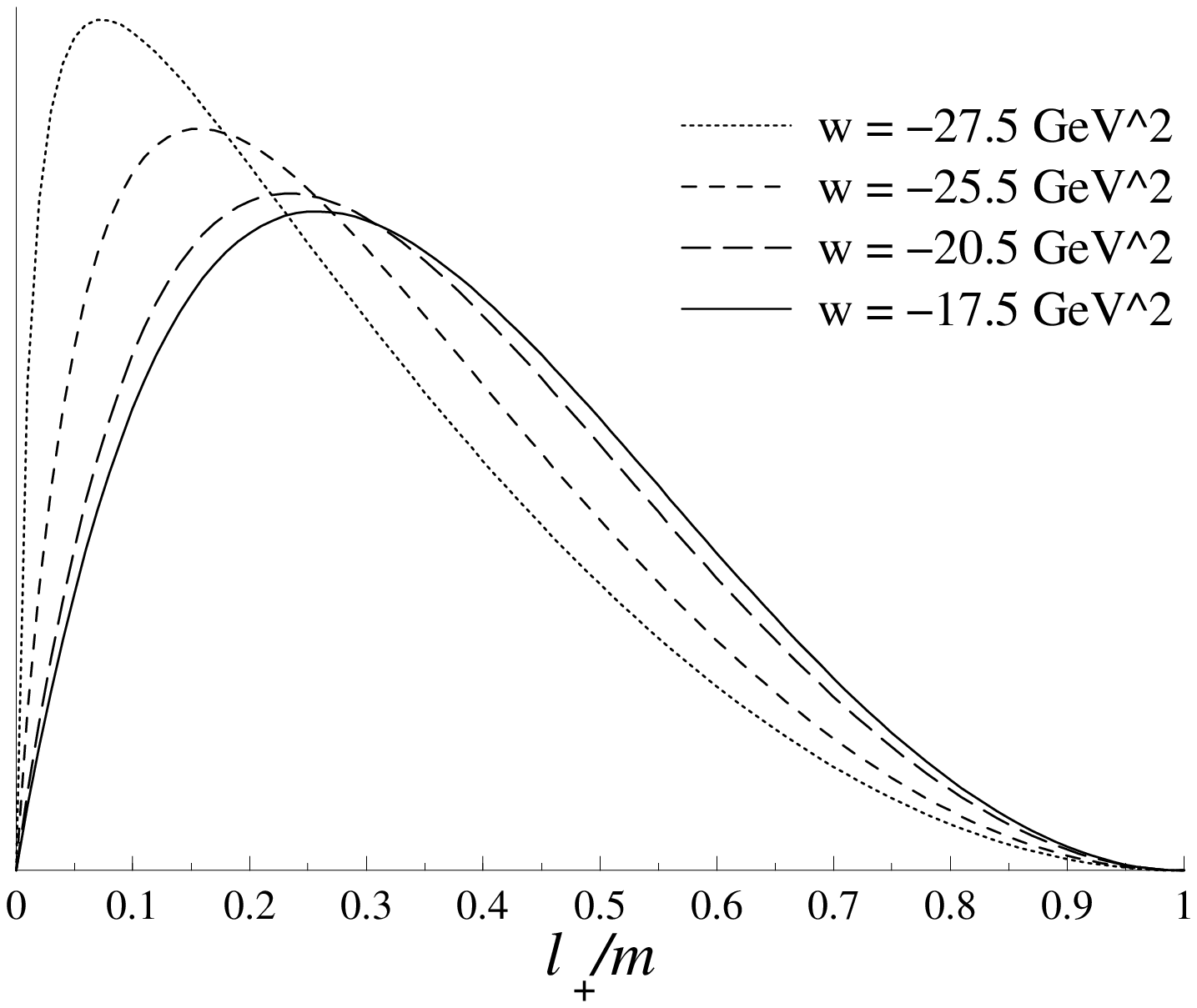}
\includegraphics[width=7cm]{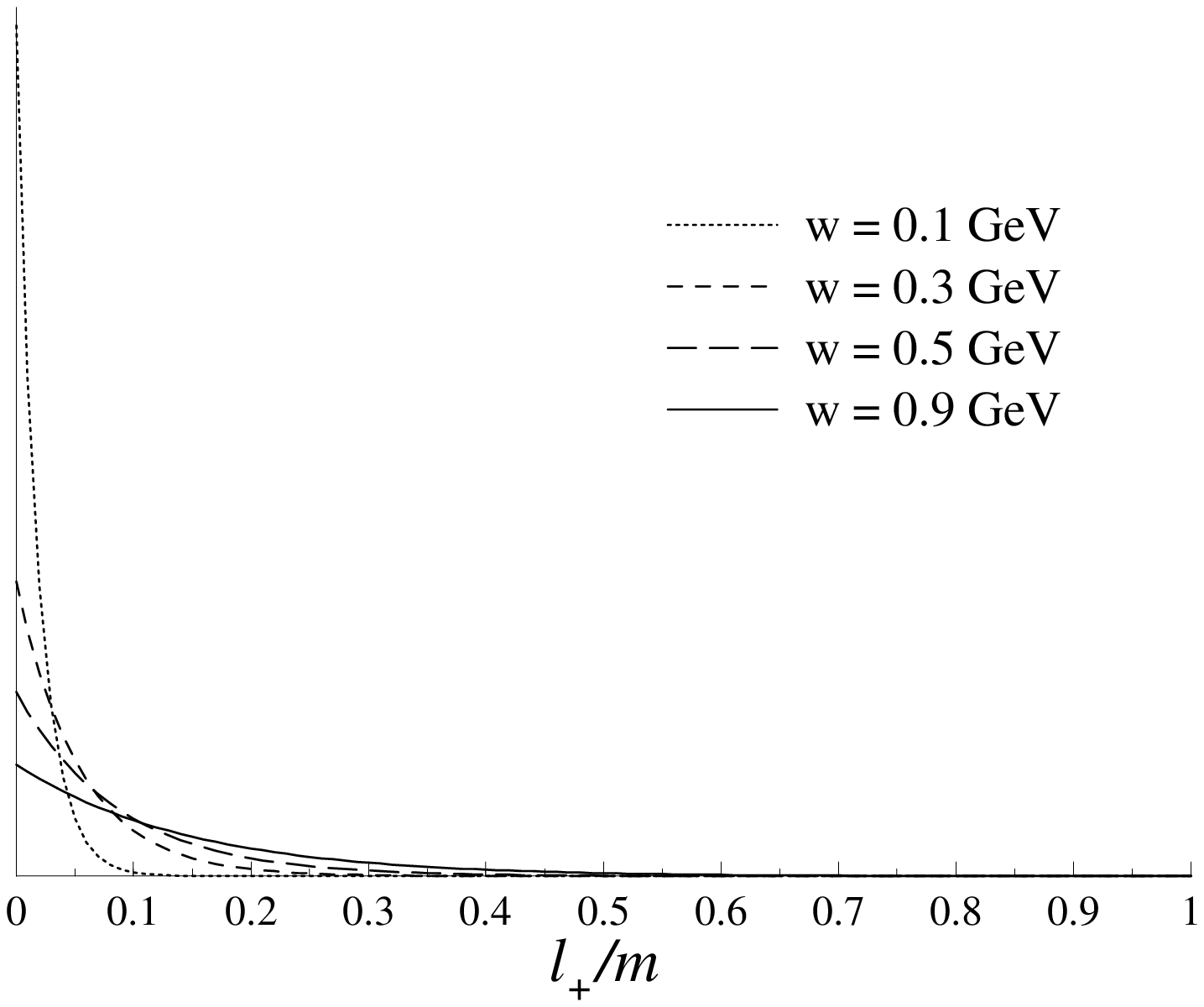}
\caption{{\footnotesize Models $\phi_{(1)}^B$, $\phi_{(2)}^B$ and
$\phi_{(3)}^B$ for the $B$ distribution amplitude, as functions
of $\xi=l_+/m$.}}
\label{fig:shapes}
\end{center}
\end{figure}

We will use three models for the ``common" $B$-meson DA; in all
three models the contribution from the region with $l_+\geq m$ is
suppressed. Defining $\xi=l_+/m$, the first two models
are~\cite{Li-Yu}:
\begin{eqnarray}
\phi_{(1)}^B(l_+)&=&N_{(1)}
  \sqrt{\xi(1-\xi)}\exp\left[-\frac{m^2}{2\omega^2}\xi^2\right]\ ,
   \label{eq:bmesmodela}\\
\phi_{(2)}^B(l_+)&=&N_{(2)}
  \frac{\xi(1-\xi)^2}{m^2+\omega(1-\xi)}\ .
   \label{eq:bmesmodelb}
\end{eqnarray}
and we introduce a third model, with a different end-point
behaviour:
\begin{equation}
\phi_{(3)}^B(l_+)=N_{(3)}
  (1-\xi)\exp\left[-\frac{m}{\omega}\xi\right]\ .
   \label{eq:bmesmodelc}
\end{equation}
The normalization constants $N_{(i)}$ are obtained from the
integral $\int dl_+ \phi^B(l_+)=1$. Fig.~\ref{fig:shapes} shows
the general shape of the three models. $l_+$ is concentrated
around $\omega/\sqrt{2}$ in the first model and we expect
$\omega=O(\lqcd)$. The second model has a broad distribution in
$l_+$, and may therefore be considered highly unlikely to be
physical. The third model is concentrated in the small $l_+$
region.

We stress that the three models in
eqs.~(\ref{eq:bmesmodela})--(\ref{eq:bmesmodelc}) are introduced
to study the dependence on the expressions
(\ref{eq:f0withsudakov}) and (\ref{eq:fpluswithsudakov}) on the
end-point behaviour of the distribution amplitudes. Our
reservations about the consistency of using a single wave function
for the $B$-meson remain of course. In particular the question of
whether $1/\lambda_B$ is finite for each of the three models
depends on whether the functions in
eqs.~(\ref{eq:bmesmodela})--(\ref{eq:bmesmodelc}) are interpreted
as corresponding to $\phi^B_+$ or $\phi^B_-$, and cannot be
answered in the approximation in which a single distribution
amplitude is used for the $B$-meson.

\subsection{Dependence on the shape}
\label{subsec:shape}

We now use eqs.~(\ref{eq:f0withsudakov}) and
(\ref{eq:fpluswithsudakov}) to compute the form factors $F_+$ and
$F_0$ for the models of the $\pi$ and $B$-meson DAs introduced in
section~\ref{subsec:modeldas}. We are able to reproduce the quoted
results in ref.~\cite{Li-Yu} for the first two models for the $B$
meson $\phi_{(1)}^B$ and $\phi_{(2)}^B$. We wish to study the
model-dependence of these results. For illustration we will
present our results for the favourable kinematic situation
$\eta=1$ (i.e. $q^2=0$, where $F_{0}=F_{+}$). The results are
presented for the following choice of parameters: $f_B=0.19$~GeV,
$m=5.28$~GeV, and $\lqcd=0.25$~GeV.

\begin{table}[ht]
\begin{center}
\begin{tabular}{|c|c|c|c|c|c|}
\hline
\multicolumn{2}{|c|}{$B$-meson DA} &
\multicolumn{4}{|c|}{Pion DA}\\
\hline
Model & $\omega$ & As. & $\alpha_2$ & $\alpha_4$ & C-Z\\
\hline
               & 0.1 & 0.28 & 0.46 & 0.46 & 0.58 \\
               & 0.2 & 0.19 & 0.35 & 0.38 & 0.42 \\
$\phi_{(1)}^B$ & 0.3 & 0.14 & 0.29 & 0.29 & 0.33 \\
               & 0.4 & 0.12 & 0.25 & 0.25 & 0.28 \\
               & 0.5 & 0.10 & 0.22 & 0.26 & 0.25 \\
               & 0.6 & 0.09 & 0.19 & 0.23 & 0.22 \\
\hline
               & -27.5 & 0.05 & 0.12 & 0.15 & 0.13 \\
               & -25.5 & 0.04 & 0.10 & 0.13 & 0.10 \\
$\phi_{(2)}^B$ & -22.5 & 0.03 & 0.09 & 0.11 & 0.09 \\
               & -20.5 & 0.03 & 0.08 & 0.11 & 0.09 \\
               & -17.5 & 0.03 & 0.08 & 0.10 & 0.08 \\
\hline
               & 0.1 & 0.36 & 0.52 & 0.51 & 0.70 \\
               & 0.3 & 0.21 & 0.36 & 0.38 & 0.45 \\
$\phi_{(3)}^B$ & 0.5 & 0.16 & 0.29 & 0.32 & 0.35 \\
           & 0.7 & 0.14 & 0.25 & 0.28 & 0.30 \\
           & 0.9 & 0.12 & 0.23 & 0.26 & 0.27 \\
\hline
\end{tabular}
\caption{{\footnotesize Dependence of $F_{0}(0)=F_{+}(0)$ on the
$B$-meson and pion DAs. The models of $B$-meson DA depend on the
shape parameter $\omega$ (in GeV for models 1 and 3,
$\textrm{GeV}^2$ for model 2). The columns ``As.'', ``$\alpha_2$''
and ``$\alpha_4$'' correspond to $\phi_\pi=6x(1-x)$,
$\phi_\pi=6x(1-x)C_2^{(3/2)}(2x-1)$ and
$\phi_\pi=6x(1-x)C_4^{(3/2)}(2x-1)$ respectively. $\phi_\pi^{(a)}$
is a linear combination of these three terms. ``C-Z'' corresponds
to the model $\phi_\pi^{(b)}$.}} \label{tab:shapedep}
\end{center}
\end{table}

For the $B$-meson DA we consider the three models introduced above
and vary the shape parameter $\omega$. We use the two models for
the pion DA in eqs.~(\ref{eq:pidawithmoments}) and
(\ref{eq:pidacz}). Eq.~(\ref{eq:f0withsudakov}) is linear in
$\phi_\pi$. If we choose the expansion in Gegenbauer polynomials
$\phi_\pi=\phi_\pi^{(a)}$, we can give separately the numerical
contributions from each of the three terms in
eq.~(\ref{eq:pidawithmoments}), i.e. the contribution from the
asymptotic DA and those proportional to $\alpha_2$ and $\alpha_4$.
The corresponding values are indicated in Tab.~\ref{tab:shapedep}.
For instance, if we choose $\phi^B=\phi_{(1)}^B$, $\omega$=0.4~GeV
and $\phi_\pi=\phi_\pi^{(a)}$, the form factors at $q^2=0$ are
$F_{0,+}(0)=0.12+\alpha_2\cdot 0.25+\alpha_4\cdot 0.25$.

We see that the computed values of $F_{0,+}(0)$ depend very
significantly on the distribution amplitudes of both mesons.
Models concentrated around small values of $l_+$ ($\phi_{(3)}^B$
for small $\omega$) or $\bar{x}$ ($\phi_\pi^{(b)}$) give larger
values than models which have a broader spread ($\phi_{(2)}^B$,
asymptotic pion DA). This can be readily understood since the
hard-scattering kernel enhances distribution end-points (without
invoking Sudakov effects the contributions from the end-point
regions are divergent).

The conclusion of our investigation is that very good control of
the behaviour of the DA's at the end-points is needed in order to
be able to compute the form factors with a precision which would
be useful for phenomenological studies. From
table~\ref{tab:shapedep} we see that with our present knowledge of
the distribution amplitudes this is not the case.

\subsection{Dependence on the cut-off in impact parameter space}
\label{subsec:impacts}

\begin{figure}[t]
\begin{center}
\includegraphics[width=7cm]{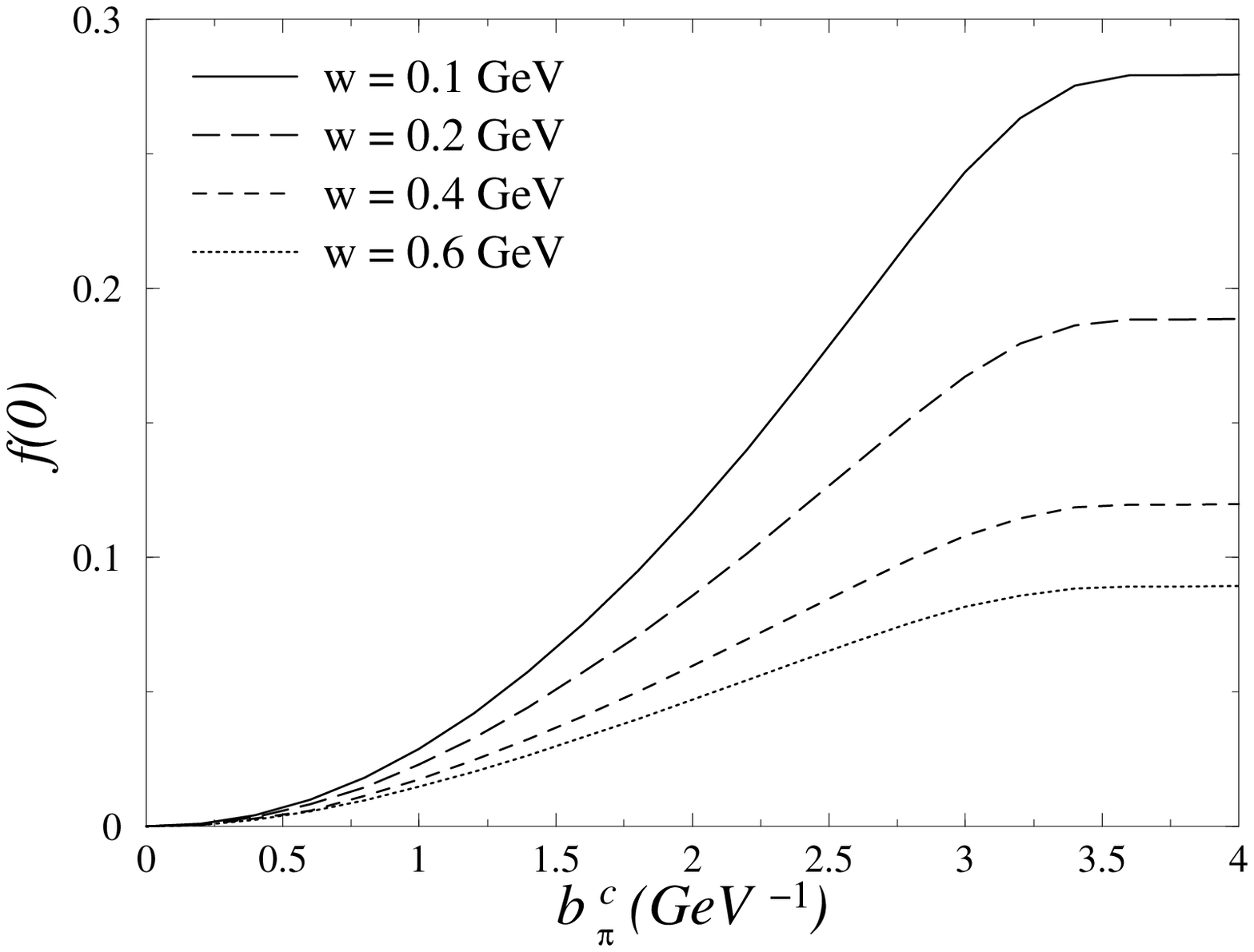}
\includegraphics[width=7cm]{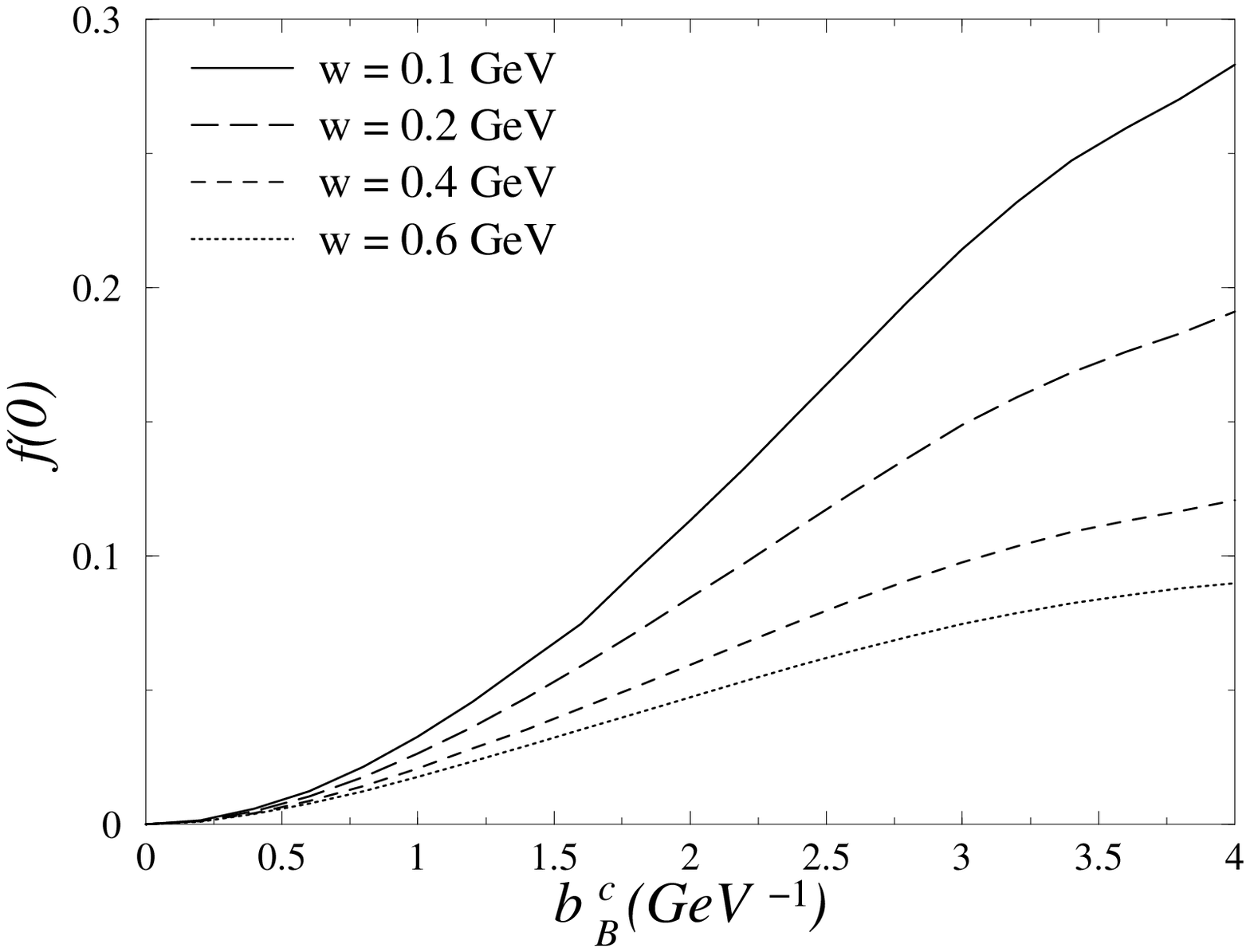}
\caption{{\footnotesize Dependence of $F_{0,+}(0)$ on the cut-offs
$b^{\,c}_\pi$ and $b^{\,c}_B$ for the integration over $\pv{b}$
(pion impact parameter) and $\pv{c}$ ($B$-meson impact parameter).
For purposes of illustration, in this figure we choose
$\phi_{(1)}^B$ as the model distribution amplitude for the
$B$-meson, and the asymptotic distribution amplitude for the
pion.}} \label{fig:cutoffs}
\end{center}
\end{figure}

In the previous subsection we have seen that with our present
knowledge of the mesons' distribution amplitudes, it is not
possible to calculate the form factors with the required
precision. We now investigate whether, for model distribution
amplitudes of the form used above, all (or almost all) of the
contribution to the form factors comes from the perturbative
region of phase space. In the preceding sections we have seen that
in order to set $\Psi_\pi(x,b_\perp,Q;\mu=1/b_\perp)$ equal to
$\phi_\pi(x;\mu=1/b_\perp)$ (see eq.~(\ref{eq:identifpsipi})\,) we
require $b_\perp$ to be small~\cite{Botts-Ster} (there is a
similar requirement for the $B$-meson). We now check whether all
of the contribution does indeed come from the region of small
impact parameters.

In fig.~\ref{fig:cutoffs} we evaluate the integrals in
eqs.~(\ref{eq:f0withsudakov}) and (\ref{eq:fpluswithsudakov}), but
with a cut-off introduced for the impact parameters for the pion
or the $B$-meson. For purposes of illustration we take
$\phi_{(1)}^B$ for the $B$-meson's DA, and the asymptotic
distribution amplitude for the pion, but similar plots can readily
be obtained for the other models of distribution amplitudes. The
figure contains plots with the integrals over the impact
parameters performed over the regions $0\leq b_\perp\leq
b^{\,c}_\pi$ and $0\leq c_\perp\leq 1/\lqcd$ (left-hand figure)
or $0\leq b_\perp\leq 1/\lqcd$ and $0\leq c_\perp\leq b^{\,c}_B$
(right-hand figure). Fig.~\ref{fig:cutoffs} shows the dependence
of $F_{0,+}(0)$ on the cut-offs $b^{\,c}_\pi$ (left) and
$b^{\,c}_B$ (right). In order for the calculations to be
consistent, we require that the contribution from the regions of
large impact parameters is negligible, and hence that the curves
in fig.~\ref{fig:cutoffs} reach a plateau for values of $b_\pi^c$
and $b_B^c$ in the perturbative region.

The purpose of our
investigation is to check whether this is the case. In general,
and for the models used in fig.~\ref{fig:cutoffs} in particular,
the answer is clearly \textit{no}. Even if we optimistically take
500~MeV as the value above which perturbation theory holds, we see
that the curves in the figure are far from saturating at this
scale. There is a significant (and uncalculable) contribution from
the nonperturbative region of phase space, where
eqs~(\ref{eq:genbotts}) and (\ref{eq:Bbotts}) are affected by
large corrections, impossible to estimate.
Contributions from regions with larger impact parameters
cannot be calculated reliably and hence we conclude that pQCD
calculations for the form factors are not valid.

A similar study was reported in ref.~\cite{Li-Yu}, in which the
authors impose that for a consistent pQCD computation, most of the
contribution should come from the region where
$\as(1/b_\perp)/\pi$ and $\as(1/c_\perp)/\pi$ are smaller than
0.5, i.e. the impact parameters are smaller than
$b^{\,c}_\pi=b^{\,c}_B
\equiv b^{\,c}\leq 0.6/\lqcd$. The authors concluded that pQCD
approach was relatively self-consistent, since a large
contribution comes from the perturbative region. Our discussion
above makes it clear that we do not accept the conclusion of
ref.~\cite{Li-Yu}.

\begin{figure}[t]
\begin{center}
\includegraphics[width=9cm]{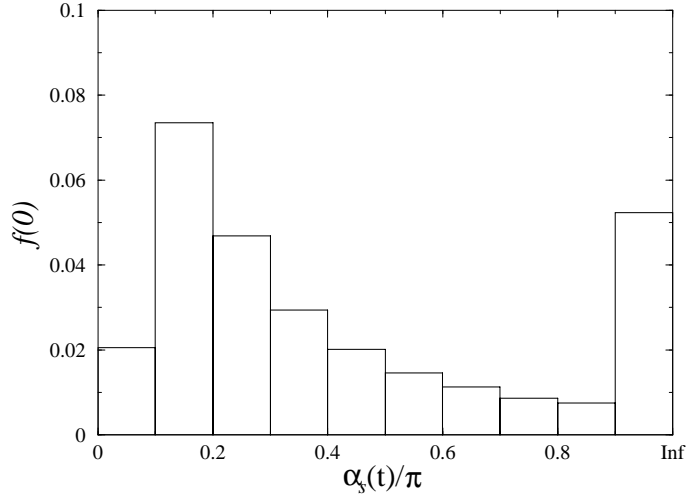}
\caption{{\footnotesize Contribution to $F_{0,+}(0)$ as a function
of the coupling constant $\as(t)$ in the hard-scattering kernel.
For purposes of illustration, in this figure we choose
$\phi_{(1)}^B$ as the model distribution amplitude for the
$B$-meson (with $\omega=0.1$ GeV), and the asymptotic distribution
amplitude for the pion.}} \label{fig:histoalphas}
\end{center}
\end{figure}

The authors of ref.~\cite{penguinpqcd} considered a weaker
criterion, that most of the contribution should come from the
region of phase space in which the maximal available virtuality
(longitudinal or transverse) of the gluon, $t\equiv
\max(\sqrt{l_+x\eta m},1/b_\perp,1/c_\perp)$ is in the
perturbative regime. They take $t$ as the scale of the coupling
constant in the hard-scattering kernel. We have argued above that
this condition is insufficient for the consistency of the
calculations, $1/b_\perp$ and $1/c_\perp$ must also separately be
in the perturbative regime. Of course, since $t\ge 1/b_\perp$ and
$t\ge 1/c_\perp$, for any set of distribution amplitudes a larger
fraction of the form-factor comes from the region is which $t>t^c$
where $t^c$ is a cut-off defining the perturbative region of phase
space, than from the regions in which $1/b_\perp>t^c$ or
$1/c_\perp>t^c$.

We have however considered this weaker criterion, and studied the
contribution to the form factors coming from the region of phase
space where $a_1\leq \as(t)/\pi \leq a_2$.
Fig.~\ref{fig:histoalphas} is the resulting histogram for
$\phi^B=\phi_{(1)}^B$, $\omega$=0.1~GeV and
$\phi_\pi=\phi_\pi^{(a)}$. The last bar on the right-hand side is
the contribution of $\as(t)/\pi$ larger than 0.9. We find that the
fraction coming from the region in which $t$ is not
perturbative, is large. Numerical studies of other models for the
distribution amplitudes indicate that the contribution from the
region of non-perturbative $t$ can be large, but this depends very
much on the choice of  distribution amplitudes. It may be
relatively small for some choices of the distribution amplitudes
\cite{penguinpqcd} and large for others.

Sudakov suppression of configurations with large transverse
separations is not efficient enough for $\btp$ form factors,
whereas it worked reasonably well in the case of the pion
electromagnetic form factor~\cite{Li-Ster}. This should not
surprise us. Recall that there is a fundamental difference between
the two processes in the BLER approach: the first suffers from
long-distance divergences, whereas the soft contributions are
finite in the latter case. One would therefore expect that a much
stronger Sudakov suppression is required, and not achieved, for a
self-consistent pQCD approach of $B$-decays.

As explained above, we have serious concerns about the derivation
and the expression of the Sudakov factor for the $B$-meson. Note
that the weakness of the Sudakov suppression in
eq.~(\ref{eq:Bbotts}) can be expected on general grounds for the
$B$-system. We have seen that a Sudakov effect was argued for
$B$-mesons with a large longitudinal light-quark momentum $l_+$,
which is a highly unlikely configuration. On the other hand, there
should be no effect for the standard situation: $l_+\sim l_-\sim
l_\perp \sim \lqcd$. ``Large'' transverse separations ($c_\perp
\geq \lqcd$) are not suppressed by this mechanism. Therefore, even
if an expression of the form (\ref{eq:Bbotts}) could be derived
for the Sudakov factor in the $B$-meson, it would lead to large
contributions coming from nonperturbative regions.

We conclude this section by restating that the pQCD predictions
for $\btp$ form factors receive a substantial (but uncalculable)
contribution from configurations with large transverse
separations. The Sudakov suppression appears to be too weak to
cure this problem. Although the numerical details of our study
depend on the choice of models for the distribution amplitudes,
the qualitative conclusion is general. We are therefore forced to
conclude that the pQCD predictions for $\btp$ semileptonic form
factors (and other physical processes which are affected by
end-point singularities) are not valid.

\section{Conclusion}\label{sec:concs}

In this paper we have studied Sudakov effects in predictions for
$\btp$ semileptonic form factors. In the standard approach, these
form-factors are not calculable due to the presence of
long-distance effects from end-point regions of (longitudinal)
phase-space. We have investigated the claim that Sudakov effects
suppress these long-distance contributions sufficiently for the
form-factors to be calculable reliably and precisely in
perturbation theory. Our conclusion is that this is not the case.
The same arguments can be used to conclude that Sudakov effects
cannot be invoked to make reliable predictions for other processes
which have end-point singularities, such as power corrections to
the amplitudes for exclusive two-body
$B$-decays~\cite{BBNS,powercorr,Li-Sanda,powcorrchin}.

Among the reasons for our conclusion are:
\begin{enumerate}
\item As explained in detail in section~\ref{sec:bdas},
it is not possible for us to accept that the formalism currently
used to derive the Sudakov factor for the $B$-meson is
theoretically sound.
\item Even if one accepts the Sudakov factors for both $B$ and
$\pi$ mesons, the uncertainty in the mesons' distribution
amplitudes (particularly at small $\bar x$ and $l_+$) means that
the form factors for $B\to\pi l\nu_l$ decays cannot be evaluated
with sufficient precision to be phenomenologically useful. This
was investigated numerically in section~\ref{subsec:shape}.
\item For these $B$-decays the kinematic parameters are such that
substantial contributions to the form-factors come from the
nonperturbative region (see sec.~\ref{subsec:impacts}), and are
hence uncalculable. Sudakov effects are too weak to suppress
contributions from the regions of phase space with large
transverse separations. We therefore conclude that the pQCD
approach is invalid for semileptonic $B$-decays.
\end{enumerate}

Following the completion of this work, there appeared the
interesting paper by H.~Kawamura et al.~\cite{kawamura}, in which
the heavy quark theory and equations of motion are used to show
that, under the assumption in which the three-parton distribution
amplitudes are neglected, the two B-meson distribution amplitudes
$\phi_+^B$ and $\phi_-^B$ can be determined (more generally it is
shown in this paper that the two- and three-parton distribution
amplitudes can be related). These results reinforce our
conclusions, in particular it is shown in ref.~\cite{kawamura}
that $\phi_-^B$ does not vanish at the end-point. Numerical
studies with these distribution amplitudes lead to the same
conclusions as presented in section~\ref{sec:numerical}.

As mentioned in the introduction, our inability to evaluate the
power corrections in two-body nonleptonic $B$-decay amplitudes
($B\to M_1 M_2$, for two mesons $M_1$ and $M_2$) severely limits
the precision with which we can deduce fundamental information
from experimental measurements of the rates and CP-asymmetries.
Our conclusion, that Sudakov suppression of long-distance effects
is too weak (and too unreliable) to be useful in extending the
range of applicability of perturbative QCD in $B$-physics, is
therefore a disappointing one. Nevertheless, given the fact that
many phenomenological studies of $B$-decays are being performed
which are based on the Sudakov suppression of end-point
singularities, we felt that it was important to articulate our
concerns about the validity and reliability of this approach.

\section*{Acknowledgements}

We thank M.~Beneke, V.~Braun, G.~Buchalla, G.~Korchemsky and H.~Li
for valuable discussions and comments.\\ This work has been partially
supported by PPARC, through grants PPA/G/O/1998/00525 and
PPA/G/S/1998/00530.

\appendix

\section{A separable model for the $B$-meson}

In this appendix we introduce a model for the two leading-twist
$B$-meson distribution amplitudes (in longitudinal and transverse
momentum space). This model is introduced for illustrative
purposes and is used in section~\ref{sec:numerical}. It satisfies
the constraints arising from the equation of motion.

\subsection{Constraints from the equations of motion.} \label{appsepmod}

We will generalise eq.~(\ref{eq:matelemb}) to include the effects
of transverse momenta. Since the hard-scattering kernel is
independent of $l_-$, we can consider eq.~(\ref{eq:matelemb}) for
$z_+=0$ (but $z^2\neq 0$). The most general decomposition is:
\begin{equation} \langle\, 0|\bar{q}_\beta(z)\, b_\alpha(0)|\bar{B}(p)\rangle
  =-\frac{if_B}{4}\left[\frac{\dirac{p}+m}{2}
     \left\{2\tilde\Psi_+^B(z^2,t)
     +\frac{\tilde\Psi_-^B(z^2,t)-\tilde\Psi_+^B(z^2,t)}{t}
        \dirac{z}\right\}
     \gamma_5\right]_{\alpha\beta}\ ,
\end{equation}
with $m=M_B=m_b$, $p=mv$ and $t=v\cdot z$ ($v=p/M_B$). A
path-ordered exponential is implicitly present in the
gauge-independent matrix element. We can introduce the Fourier
transforms:
\begin{equation}
\Psi^B_\pm(l_+,\pv{l})=
  \int \frac{d^2\pv{z}}{(2\pi)^2} \frac{dz_-}{2\pi}
    e^{i(l_+z_--\pv{l}\cdot\pv{z})} \tilde\Psi_\pm^B(z_-,-{\pv{z}\,}^2)\ .
\end{equation}
We have defined $l=(l_+/\sqrt{2},0,\pv{l})$ and $z=(0,z_-\sqrt{2},\pv{z})$,
so that $z^2=-\pv{z}^{\,2}$ and $t=z_-$.

In ref.~\cite{Ben-Fel} constraints were derived on the
leading-twist $B$-meson DAs, neglecting the effects of
three-particle ($q\bar{q}g$) and higher Fock states:
\begin{eqnarray}
\left.\frac{\partial\tilde\Psi_-^B}{\partial t}
  +\frac{\tilde\Psi_-^B-\tilde\Psi_+^B}{t}
 \right|_{z^2=0} &=& 0\ ,\\
\left.\frac{\partial\tilde\Psi_+^B}{\partial z^2}
     +\frac{1}{4}\frac{\partial^2\tilde\Psi_-^B}{\partial t^2}
 \right|_{z^2=0} &=& 0\ ,
\end{eqnarray}
where $\tilde\Psi_\pm^B$ depend on $z^2$ and $t=v\cdot z$. We
consider a model in which the dependence on the longitudinal and
transverse momenta is factorized:
\begin{equation}\label{eq:separable}
\Psi^B_\pm(l_+,\pv{l})=\phi^B_\pm(l_+)\times \tau^B_\pm(l_\perp),
\end{equation}
where we chose the normalization conditions
\begin{equation}
\int dl_+\phi^B_\pm(l_+)=1\ \ \ \ \textrm{and}\ \ \ \ \int
d^2\pv{l}\tau_\pm(l_\perp)=1\ .
\end{equation}

In this case, the two constraints become in the momentum space:
\begin{eqnarray}
\phi^B_+(l_+)&=&-l_+ \frac{d\phi^B_-}{dl_+}(l_+)\ ,\\ l_+^2
\phi^B_-(l_+)&=&\lambda^2 \phi^B_+(l_+)\ ,
\end{eqnarray}
where $\lambda^2=\int d^2\pv{l} \pv{l}^2 \tau_+(l_\perp)$.
Combining the two equations, we obtain the differential equation:
\begin{equation}
\frac{d\phi^B_-}{dl_+}(l_+)=-\frac{l_+}{\lambda^2} \phi_-^B(l_+)\
,
\end{equation}
and the corresponding (normalized) solutions:
\begin{equation}
\phi_-^B(l_+)=\sqrt\frac{2}{\pi \lambda^2}
  \exp\left[-\frac{l_+^2}{2 \lambda^2}\right]
\qquad\textrm{and}\qquad  \phi_+^B(l_+)=\sqrt{\frac{2}{\pi
\lambda^2}}
  \frac{l_+^2}{\lambda^2}
  \exp\left[-\frac{l_+^2}{2 \lambda^2}\right]\ .
\end{equation}

The value of $\lambda$ depends on the model used for the
transverse momentum, and measures the dispersion of its values.
For example, the step-function distribution:
\begin{equation}\label{eq:flat}
\tau_+(l_\perp)=\left\{
\begin{array}{cl}
1/(\pi \lambda_\perp^2) & {\textrm{for}}\
  0\leq l_\perp\leq \lambda_\perp\\
0 & {\textrm{otherwise}}
\end{array}
\right.
\end{equation}
leads to $\lambda=\lambda_\perp/\sqrt{2}$, whereas the Gaussian
distribution:
\begin{equation}
\tau_+(l_\perp)=\frac{1}{2\pi\sigma_\perp^2}
   \exp\left[-\frac{l_\perp^2}{2\sigma_\perp^2}\right]
\end{equation}
has $\lambda=\sqrt{2}\sigma_\perp$.

\subsection{Can we set $\Psi_+^B=\Psi_-^B$?} \label{sec:errest}

As an exploratory exercise we estimate the error in the calculated
value of the form-factors (see eqs.~(\ref{eq:f0p}),
(\ref{eq:eqi0}) and (\ref{eq:eqip})) caused by setting $\Psi_+^B$
equal to $\Psi_-^B$. For this exercise, we take for the DA of the
$B$-meson the separable model of App.~\ref{appsepmod}, with a flat
transverse distribution, eqs.~({\ref{eq:separable}) and
(\ref{eq:flat}). We set $\lambda_\perp=\Lambda$, with $\Lambda=O(\lqcd)$.
For this model the
first term in eq.~(\ref{eq:eqi0}) is:
\begin{eqnarray}
S_-(\eta)&=&\int_0^1 dx \phi_\pi(x)
  \int_0^\infty dl_+ \phi^B_-(l_+) \int d^2\pv{k} \tau_+(k_\perp)
   \int d^2\pv{l} \tau_+(l_\perp)\\
 &&\qquad \times\eta \frac{1}{\bar{x} l_+ \eta m + (\pv{k}+\pv{l})^2}
    \frac{1}{\bar{x}\eta m^2+\pv{k}^2}\nono\\
 &=& \eta \left(\frac{2\pi}{\pi\Lambda^2}\right)^2
      \int_0^1 dx \phi_\pi(x) \int_0^\infty dl_+ \phi^B_-(l_+)
      \int_0^\Lambda k_\perp dk_\perp
   \int_0^\Lambda l_\perp dl_\perp\\
 &&\qquad
   \times\frac{1}{\bar{x}\eta m^2+\pv{k}^2}
   \frac{1}
 {\sqrt{(\bar{x} l_+ \eta m + k_\perp^2 + l_\perp^2)^2-4k_\perp^2
   l_\perp^2}}
 \nono\\
 &=& \frac{\eta}{\Lambda^4} \int_0^1 dx \phi_\pi(x)
 \int_0^\infty dl_+ \phi^B_-(l_+)
 \int_0^{\Lambda^2} d\kappa \int_0^{\Lambda^2} d\lambda
   \label{eq:mmint}\\
 &&\qquad
\times\frac{1}{\bar{x}\eta m^2+\kappa}
   \frac{1}
   {\sqrt{(\bar{x} l_+ \eta m + \kappa + \lambda)^2-4\kappa\lambda}}
   \ .\nono
\end{eqnarray}
The following integral is useful for the angular integration over
the transverse momenta:
\begin{equation}
\int_0^{2\pi} \frac{1}{a+b\cos\theta} d\theta=
   \frac{2\pi}{\sqrt{a^2-b^2}}
 \qquad \textrm{for}\ a>b\ .
\end{equation}
In eq.~(\ref{eq:mmint}), the integrals over $\lambda$, $\kappa$
and $l_+$ can be performed numerically and we find :
\begin{eqnarray}
I_-(\bar{x},\eta)&=&\frac{\eta}{\Lambda^4}\int_0^\infty dl_+
 \phi^B_-(l_+)
    \int_0^{\Lambda^2} d\kappa \int_0^{\Lambda^2} d\lambda\\
 &&\qquad      \times\frac{1}{\bar{x}\eta m^2+\kappa}
\frac{1}{\sqrt{(\bar{x} l_+ \eta m + \kappa + \lambda)^2
    -4\kappa\lambda}}\ .
   \nono
\end{eqnarray}
The same integral can be computed with $\Psi_+^B$ instead of
$\Psi_-^B$, and the result is called $I_+(\bar{x},\eta)$. For
illustration, the two functions are plotted in
Fig.~\ref{fig:mymod} for the case $\eta=1$. Throughout the region
in $x$, $I_-$ remains about twice as large as $I_+$.

\begin{figure}[t]
\begin{center}
\includegraphics[width=11cm]{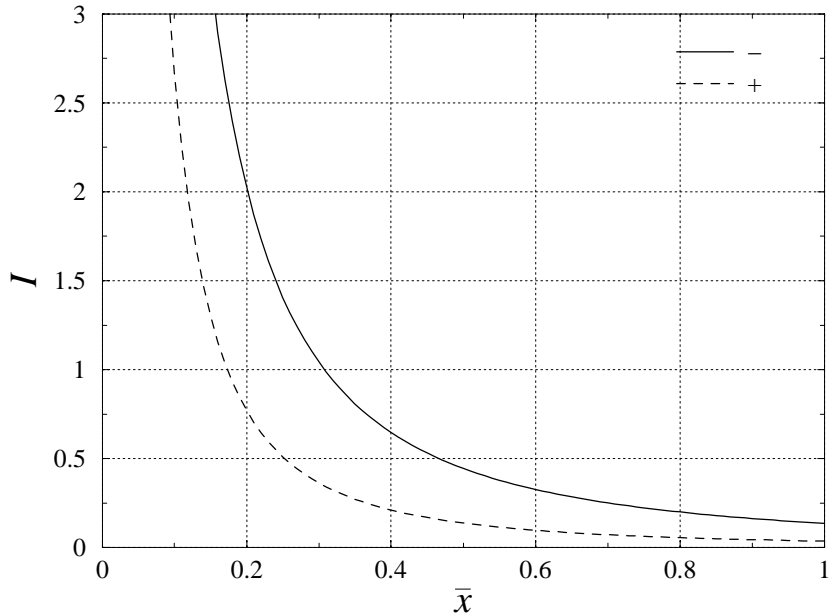}
\caption{{\footnotesize First term
$I_-(\bar{x},\eta=1)$ in eq.~(\ref{eq:eqi0}), with
the separable model for $\Psi_-^B$, and corresponding integral
$I_+(\bar{x},\eta=1)$ when
$\Psi_-^B$ is replaced by $\Psi_+^B$.}}
\label{fig:mymod}
\end{center}
\end{figure}

In order to estimate the error on the form factors we need to take
a model for the DA of the pion. For example, if we take the
asymptotic form, $\phi_\pi(x)=6x(1-x)$, we obtain
$S_-(\eta=1)=1.32~\textrm{GeV}^{-4}$ and
$S_+(\eta=1)=0.59~\textrm{GeV}^{-4}$. The identification of $\Psi_+$
and $\Psi_-$ leads therefore to an error of at least 30\% in this
comtribution to the form factors.

\end{document}